\newcommand{\blind}{1}
\documentclass[authoryear,12pt]{article}
\usepackage[a4paper, total={6.5in, 10in}]{geometry}
\usepackage{setspace}
\usepackage{natbib}
\usepackage[export]{adjustbox}
\usepackage{import}
\usepackage{xifthen}
\usepackage{pdfpages}
\usepackage{transparent}
\usepackage{fontenc, inputenc, subcaption, graphicx, sectsty, titlesec, titling, caption, fancyhdr}
\usepackage{amsthm, amsmath, amssymb, nccmath, xcolor}
\usepackage{paralist, multirow, arydshln}
\usepackage[noend]{algpseudocode}
\usepackage[ruled, vlined]{algorithm2e}
\usepackage{float}
\usepackage[colorlinks=true,linkcolor=black, citecolor=blue, urlcolor=purple]{hyperref}
\usepackage{comment}

\def\0{\mbox{\boldmath{$\mathbf{0}$}}}
\def\1{\mbox{\boldmath{$\mathbf{1}$}}}
\def\bzeta{\mbox{\boldmath$\zeta$}}

\def\btheta{\mbox{\boldmath$\theta$}}

\def\bmu{\mbox{\boldmath$\mu$}}
\def\bphi{\mbox{\boldmath$\phi$}}
\def\bbeta{\mbox{\boldmath$\beta$}}

\def\bzeta{\mbox{\boldmath$\zeta$}}

\def\bgamma{\mbox{\boldmath$\gamma$}}
\def\bGamma{\mbox{\boldmath$\Gamma$}}
\def\bsigma{\mbox{\boldmath$\sigma$}}

\def\blambda{\mbox{\boldmath$\lambda$}}

\def\A{\mbox{\boldmath{$\mathbf{A}$}}}

\def\F{\mbox{\boldmath{$\mathbf{F}$}}}

\def\I{\mbox{\boldmath{$\mathbf{I}$}}}

\def\R{\mbox{\boldmath{$\mathbf{R}$}}}

\def\X{\mbox{\boldmath{$\mathbf{X}$}}}

\def\Z{\mbox{\boldmath{$\mathbf{Z}$}}}

\def\c{\mbox{\boldmath{$\mathbf{c}$}}}

\def\f{\mbox{\boldmath{$\mathbf{f}$}}}

\def\s{\mbox{\boldmath{$\mathbf{s}$}}}

\def\w{\mbox{\boldmath{$\mathbf{w}$}}}
\def\x{\mbox{\boldmath{$\mathbf{x}$}}}
\def\y{\mbox{\boldmath{$\mathbf{y}$}}}
\def\z{\mbox{\boldmath{$\mathbf{z}$}}}

\newcommand{\trans}{^{T}} 

\begin{document}
	\def\spacingset#1{\renewcommand{\baselinestretch}%
		{#1}\small\normalsize} \spacingset{1}
	
	\if1\blind
	{
		\title{\bf Bayesian Variable Selection in Double Generalized Linear Tweedie Spatial Process Models}
		\author{Aritra Halder$^{a,*\dagger}$, Shariq Mohammed$^{b,c\dagger}$ and Dipak K. Dey$^{d}$\\~\\
			$^{a}${\small Department of Biostatistics, Dornsife School of Public Health,} \\{\small Drexel University, Philadelphia, PA, USA}\\
			$^{b}${\small Department of Biostatistics, Boston University School of Public Health,} \\{\small Boston University, Boston, MA, USA}\\
                $^{c}${\small Rafik B. Hariri Institute for Computing and Computational Science \& Engineering,} \\{\small Boston University, Boston, MA, USA}\\
			$^{d}${\small Department of Statistics, University of Connecticut, Storrs, CT, USA}\\
			$^{*}${\small corresponding author. E-mail: aritra.halder@drexel.edu}\\
			$^{\dagger}${\small equal contribution}
		}
		\maketitle
	} \fi
	
	\if0\blind
	{
		\bigskip
		\bigskip
		\begin{center}
			{\Large\bf Bayesian Variable Selection in Double Generalized Linear Tweedie Spatial Process Models}
		\end{center}
		\medskip
	} \fi
	
	\bigskip
	\begin{abstract}
		Double generalized linear models provide a flexible framework for modeling data by allowing the mean and the dispersion to vary across observations. Common members of the exponential dispersion family including Gaussian, compound Poisson-gamma, Gamma, inverse-Gaussian are known to admit such models. However, the lack of their use can be attributed to ambiguities that exist in model specification under a large number of covariates and, complications that arise when data from a chosen application display complex spatial dependence. In this work we consider a hierarchical specification for these models with a spatial random effect. The spatial effect is targeted at performing uncertainty quantification by modeling dependence within the data arising from location based indexing of the response. We focus on a Gaussian process specification for the spatial effect. Simultaneously, we tackle the problem of model specification for such hierarchical spatial process models using Bayesian variable selection. It is effected through a continuous spike and slab prior placed on the model parameters (or fixed effects). The novelty of our contribution lies in the Bayesian frameworks developed for such models, which have not been explored previously. We perform various synthetic experiments to showcase the accuracy of our frameworks. These developed frameworks are then applied to analyse automobile insurance premiums in Connecticut.
	\end{abstract}
	
	\noindent%
	\vfill
	
	\newpage
	\spacingset{1.45} 
	
\section{Introduction}\label{sec: intro}
Spatial processes have occupied center stage in statistical theory and applications for the last few decades. Their voracious use can largely be explained by geographically tagged data becoming increasingly commonplace in modern applications. Such data are often composed of complex variables which are no longer amenable to a Gaussian assumption. For example, spatially indexed counts \citep[see for e.g.,][]{wolpert1998poisson, best2000spatial,agarwal2002zero,lawson2018bayesian}, proportions \citep[see for e.g.,][]{gelfand2000modeling, gelfand2005modelling, finley2009hierarchical, eidsvik2012approximate}, time to event or, survival outcomes \citep[see for e.g.][]{banerjee2004parametric, martino2011approximate, zhou2015bayesian} are some frequently occurring variables where spatial processes have proved invaluable in performing uncertainty quantification. The purpose being to quantify unobserved dependence introduced within the variable of interest due to varying location. The cornerstone for such studies is a spatially indexed process variable of interest, often termed as a response process and denoted by, $y(\s)$. This is accompanied by covariate information $\X(\s)=[\x_1(\s),\x_2(\s),\ldots,\x_p(\s)]$. Here $\s\in {\cal S} = \{\s_1,\s_2,\ldots,\s_L\}$ is the spatial indexing and, ${\cal S}$ is a finite set of indices or, locations over which the response variable and covariates are observed. The investigator often encounters observation--level covariates that account for response specific characteristics when learning such processes. It becomes important to understand which of these covariates are important contributors to variation in the data. From a model parsimony standpoint, model choice becomes an important issue to the investigator. In statistical theory, this problem is often addressed by performing shrinkage or, variable selection on the model coefficients. Moreover, performing spatial uncertainty quantification produces accurate inference for model coefficients, which also raises the concern regarding a more ``honest" subset of covariates within $\X(\s)$ that primarily determine the variation in $y(\s)$. The crown jewel of our contribution is Bayesian methodology for performing spatial uncertainty quantification and model choice simultaneously.

Spatial process modeling generally requires a hierarchical specification of an unobserved random effect within the model \citep[][]{clark2006future}. Maintaining a hierarchical approach allows for exploration of the effect of covariates, $\X(\s)$ and, the random effect $\w(\s)$ jointly on the response process $y(\s)$. Particularly, considering generalized linear spatial process modeling, it is assumed that $y(\s)\mid\bbeta,\w(\s)$ arise in a {\em conditionally} independent fashion from a member of the exponential family with mean $\mu(\s)$ such that $g(\mu(\s))=\X(\s)\bbeta+\F(\s)\w(\s)$, where $g$ is a monotonic link function, $\bbeta$ are model coefficients, or fixed effects and $\F(\s)$ is a spatial incidence matrix. In contrast to Gaussian response processes where a direct hierarchical specification on the response is feasible, modeling a non-Gaussian spatial process leverages the generalized linear model framework to employ a latent process specification \citep[see for e.g.,][]{zeger1991generalized, diggle1998model, dey2000generalized, zhang2002estimation, banerjee2014hierarchical}. This is facilitated by the existence of a valid joint probability distribution, $\pi\left(y(\s_1),y(\s_2),\ldots,y(\s_L)\mid\bbeta,\btheta_{pr}\right)$, where $\btheta_{pr}$ denotes process parameters required for specification of $\w(\s)$ \citep[see discussion in section 6.2,][]{banerjee2014hierarchical}. This hierarchical specification gives us a natural way to perform variable selection by incorporating shrinkage priors into the hierarchical prior formulation. There are several choices of shrinkage priors that differ in their prior specification such as: the discrete spike-and-slab prior \citep[see, for e.g,][]{mitchell1988bayesian,george1993variable}, other priors based on the Gaussian-gamma family in linear Gaussian models \citep[see, for e.g,][]{raftery1997bayesian,berger2001objective}, the continuous spike-and-slab prior \citep[see, for e.g,][]{ishwaran2005spike}, the Bayesian counterparts of LASSO and elastic net \citep[see, for e.g,][]{park2008bayesian,li2010bayesian}, mixtures of $g$--priors \citep[see, for e.g,][]{liang2008mixtures}, the horseshoe priors and its variants \citep[see, for e.g,][]{carvalho2010horseshoe}, among several others. We use the continuous spike-and-slab prior to effect shrinkage on model coefficients.

 We focus on a subset of probability distributions within the exponential family, termed as the exponential dispersion family \citep[see, for e.g.,][]{jorgensen1986some,jorgensen1987exponential,jorgensen1992exponential,jorgensen1997theory}. It allows the dispersion along with the mean to vary across observations, suppressing the need for having a constant dispersion across observations. We focus on a particular member of the family, the Tweedie compound Poisson-gamma (CP-g), more commonly referred to as the Tweedie (probability) distribution \citep[see,][]{tweedie1984index}. The corresponding random variable is constructively defined as a Poisson sum of independently distributed Gamma random variables. Allowing for a varying dispersion across observations enables exploration of the effect of covariates $\X(\s)$ on the mean and the dispersion separately, by employing two separate generalized linear models (GLMs). This gives rise to the double generalized linear model (DGLM) \citep[see, for e.g.,][]{smyth1989generalized,verbyla1993modelling,smyth1999adjusted}. Hierarchical frameworks for specifications of DGLMs were first developed in \cite{lee2006double}. Although not mandatory, it is customary to use the same covariates, $\X(\s)$ for both the mean and dispersion GLMs to avoid ambiguities in model specification. Previous work \citep[see for e.g.,][]{halder2022spatial, halder2019spatial} uses this approach and considers developing inference for DGLMs under a frequentist framework. Inference on spatial effects is obtained through penalizing the graph Laplacian. In this paper we adopt a Bayesian discourse by supplementing the DGLM framework with the continuous version of the spike-and-slab prior to effect shrinkage and thereby achieve better model specification. We integrate the spike and slab prior into our hierarchical prior formulation for both mean and dispersion models. We show that these priors provide a natural way of incorporating sparsity into the model, while offering straightforward posterior sampling in the context of our spatial DGLMs. 
	
 The scale for spatial indexing is assumed to be point-referenced. For example, latitude-longitude or easting--northing. Generally, $\s \in \mathbb{R}^2$. Specification of a neighborhood structure or, proximity is naturally important when attempting to quantifying the behavior of response in locations that are {\em near} each other. We select the Euclidean distance between locations. This results in a Gaussian process \citep[see, for e.g.,][]{williams2006gaussian} prior on the spatial process, $\w(\s)$.  Other choices exist for specifying such spatial process, $\w(\s)$, for e.g. log-Gamma \citep[see][]{bradley2018computationally,bradley2020bayesian} etc. We particularly focus on a Gaussian process specification on $\w(\s) \sim GP(\0,K(\cdot))$, where $K$ is a covariance function. For arbitrary locations, $\s$ and $\s'$, dependence between $y(\s)$ and $y(\s')$ is specified through $K(\s,\s')$, which governs the covariance between $w(\s)$ and $w(\s')$. For point-referenced data, the Mat\'ern family \citep[see, for e.g.,][]{matern2013spatial} provides the most generic and widely adopted covariance specification.
	
 Next, we address Bayesian model specification. In the absence of such concerns for the hierarchical process models discussed above, prior specification follows the generic framework, $[\text{data}\mid\text{process}, \widetilde{\btheta}]\times [\text{process}\mid\widetilde{\btheta}]\times [\btheta_{m}]\times [\btheta_{pr}]$. Here, $\widetilde{\btheta}=\{\btheta_{m}$, $\btheta_{pr}\}$ denote model parameters \citep[see for e.g.][Chapter 6, p. 125]{berliner2000hierarchical, gelfand2017bayesian, banerjee2014hierarchical}. In particular, $\btheta_{pr}$ constitute parameters instrumental in specification of the process, while $\btheta_m$ are other model parameters. We adopt a proper prior on $\btheta_{pr}$ to avoid the risk of generating improper posteriors \citep[see, for e.g.,][]{berger2001objective1}. Building a Bayesian variable selection framework that facilitates model specification for $\btheta_{m}$ requires an additional layer of hierarchical prior specification, appending the latter framework with variable selection parameters, $\btheta_{vs}$ and, thereby producing
	\begin{equation}\label{eq:modelform}
	    [\text{data}\mid\text{process}, \btheta]\times [\text{process}\mid\btheta]\times [\btheta_{pr}]\times [\btheta_{m}]\times [\btheta_{vs}],
	\end{equation}
 where $\btheta=\{\widetilde{\btheta}, \btheta_{vs}\}$. We resort to Markov Chain Monte Carlo (MCMC) sampling \citep[see, for e.g.][]{carlin2008bayesian,girolami2011riemann} for performing posterior inference on $\btheta$. The novelty of our approach lies in the simple Bayesian computation devised---employing only Gibbs sampling updates for $\btheta_{vs}$. To the best of our knowledge {\em hierarchical Bayesian frameworks} for fitting (a) Tweedie DGLMs, (b) spatial Tweedie DGLMs with (or without) variable selection, do not exist in the statistical literature. Evidently, proposed methodology in (\ref{eq:modelform}) remedies that. 
	
 The ensuing developments in the paper are organized as follows: In Section \ref{sec:sf} we detail the proposed statistical framework outlining Tweedie distributions---the likelihood and parameterization, model formulation and the hierarchical prior specification. Section \ref{sec:synexp} provides comprehensive synthetic experiments that document the efficacy of our proposed statistical framework for Bayesian variable selection in spatial DGLMs. Section \ref{sec:app} considers application of the developed framework to automobile insurance premiums in Connecticut, USA during 2008. Additional synthetic experiments capturing various performance aspects for the models are provided in the Supplementary Material. 

\section{Statistical Framework}\label{sec:sf}
The Tweedie distribution produces observations composed of exact zeros with a continuous Gamma tail. Their ability to model mixed data types featuring exact zeros and continuous measurements jointly makes them suitable for modeling response arising from a variety of domains. Some of the current applications include, actuarial science \citep[see][]{smyth2002fitting, yang2018insurance, halder2022spatial, halder2019spatial}, ecology \cite[see for e.g.,][]{swallow2016bayesian}, public health \citep{ye2021comparisons}, environment \citep[see for e.g.,][]{bonat2021tweedie}, ecology \citep[][]{shono2008application}, gene expression studies \citep{mallick2022differential}. As evidenced by these applications, the presence of unobserved dependence between observations, affecting the quality of inference, is not unlikely. In the following subsections we provide more details on Tweedie distributions, followed by the model formulation and hierarchical prior specification.

\subsection{The Exponential Dispersion Family: Tweedie Distributions}
The Tweedie family of distributions belong to the exponential dispersion (ED) family of models whose probability density/mass function has the generic form,
	\begin{equation}\label{eq:ed}
	    \pi(y\mid\theta,\phi) = a(y,\phi)\exp\left\{\phi^{-1}(y\theta-\kappa(\theta))\right\},
	\end{equation}
	where $\theta$ is the natural or canonical parameter, $\phi$ is the dispersion parameter, and $\kappa(\theta)$ is the cumulant function. Characterizing the Tweedie family is an index parameter $\xi$, varying values of which produce different members of the family. 
    For e.g., the CP-g is obtained with $\xi\in (1,2)$, for $\xi = 1$ we obtain a Poisson and $\xi = 2$ produces a Gamma distribution, for $\xi\in (0,1)$ they do not exist \citep[for further details see Table 1,][]{halder2022spatial}. We are particularly interested in the CP-g distributions in this work. In general, for the ED family we have the mean, $\mu = E(y)=\kappa'(\theta)$ and the variance, $Var(y)=\phi\kappa''(\theta)$. For the CP-g we have $\kappa(\theta)=(2-\xi)^{-1}\{(1-\xi)\theta\}^{2-\xi/1-\xi}$. Using the relation, $\kappa'(\theta) = \mu$, some straightforward algebra yields, $\kappa(\theta) = (2-\xi)^{-1}\mu^{2-\xi}$ and $\kappa''(\theta)=\mu^\xi$, implying $Var(y)=\phi\mu^\xi$ and denoting $\alpha=(1-\xi)^{-1}(2-\xi)$ we have,
	\begin{equation*}
            a(y,\phi)= 1\cdot I(y=0) + y^{-1}\sum_{j=1}^{\infty}\left[\frac{y^{-\alpha}(\xi-1)^\alpha}{\phi^{1-\alpha}(2-\xi)}\right]^{j}\frac{1}{j!\Gamma(-j\alpha)}I(y>0).
	\end{equation*}
	Evidently, $\pi(0\mid\theta, \phi) = \exp\{-\phi^{-1}\kappa(\theta)\}$.
 
 We introduce some notation. Let $y_{ij}(\s_i)$ denote the $j$-th response at the $i$-th location $\s_i\in {\cal S}$, where $j = 1, 2,\ldots, n_i$ and $i = 1, 2,\ldots, L$ with $\sum_{i=1}^{L}n_i=N$. Together we denote, $\y=\y(\s)=\{\{y_{ij}(\s_i)\}_{j=1}^{n_i}\}_{i=1}^{L}$ as the $N\times 1$ response. Similarly, $\bmu=\bmu(\s) = \{\{\mu_{ij}(\s_i)\}_{j=1}^{n_i}\}_{i=1}^{L}$ and $\bphi= \{\{\phi_{ij}\}_{j=1}^{n_i}\}_{i=1}^{L}$ denotes the mean and dispersion vectors respectively. If $\y\mid\bmu, \bphi,\xi$ arises independently from a CP-g distribution, then the likelihood is given by
    \begin{equation}\label{eq:likelihood}
        \pi(\y\mid\bmu,\bphi,\xi) = \prod_{i=1}^{L}\prod_{j=1}^{n_i}a_{ij}(y_{ij}(\s_i)\mid\phi_{ij})\times\exp\left[\phi_{ij}^{-1}\left(\frac{y_{ij}(\s_i)\mu_{ij}(\s_i)^{1-\xi}}{1-\xi}-\frac{\mu_{ij}(\s_i)^{2-\xi}}{2-\xi}\right)\right].
    \end{equation}
	Working with the likelihood, $\pi(\cdot)$, when devising computation, evaluating the infinite series representation of $a(y,\phi)$ is required. The two commonly used methods are---saddle-point approximation \citep[see for e.g.][]{nelder1987extended,smyth2002fitting, dunn2005series, zhang2013likelihood} and Fourier inversion \citep[see, for e.g.][]{dunn2008evaluation}. The saddle-point approximation to (\ref{eq:ed}) uses a deviance function based representation where, $\widetilde{\pi}(y\mid\mu,\phi) = b(y,\phi)\exp\left\{-(2\phi)^{-1}d(y\mid\mu)\right\}$. For CP-g distributions, the deviance function is $d(y\mid\mu)=d(y\mid\mu,\xi) = 2\left\{\left(y^{2-\xi}-y\mu^{1-\xi}\right)(1-\xi)^{-1}-\left(y^{2-\xi}-\mu^{2-\xi}\right)(2-\xi)^{-1}\right\}$, and $b(y\mid\phi,\xi)=(2\pi \phi y^\xi)^{-1/2}I(y>0)+1\cdot I(y=0)\approx a(y,\phi)\exp\left\{\phi^{-1}y^{2-\xi}(1-\xi)^{-1}(2-\xi)^{-1}\right\}$. We performed experiments which showed that the saddle-point approximation performs well when fewer zeros are present in the data. Under higher proportions of zeros its performance was sub-optimal. However, albeit its computationally intensive nature, in all scenarios the Fourier inversion based method had stable performance. Hence, in this paper we use the evaluation of $a(y,\phi)$ that is based on Fourier inversion. The adopted Bayesian approach requires MCMC computation that relies on accurate likelihood evaluations. Hence, we emphasize the importance of choosing the appropriate likelihood function for application purposes. 
    We denote the likelihood in (\ref{eq:likelihood}) as $Tw(\bmu,\bphi,\xi)$. Tweedie distributions are the only members of the ED family that possess a scale invariance property \citep[see, for e.g.,][Theorem 4.1]{jorgensen1997theory}. This suggests for $c_{ij}>0$, $\y^{*}(\s)= \{c_{ij} y_{ij}(\s_i)\}\sim Tw(\c\trans\bmu,{\c^{2-\xi}}\trans\bphi,\xi)$ allowing observations with different scales of measurement to be modeled jointly.

    \subsection{Model Formulation}
    Formulating DGLMs with spatial effects theoretically involves specification of a spatial random effect in both mean and dispersion models. In such a scenario complex dependencies can be specified to account for varied degrees of uncertainty quantification. In the simplest case the corresponding spatial random effects for the mean and dispersion models are independent Gaussian processes. More complex scenarios can feature dependent Gaussian processes, where the dependence arises from a cross-covariance matrix. Spatial random effects in the mean model are readily interpretable---risk faced (adjustment to mean premium paid in our case) owing to location. However, spatial random effects in the dispersion model are not readily interpretable. Subsequently, we choose to include spatial random effects only in the mean model for this work.
    
    Let $\x_{ij}(\s_i)$ denote a $p\times 1$ vector of observed covariates for the mean model and $\bbeta$ be the corresponding a $p\times 1$ vector of coefficients. $\f_{ij}(\s_i)$ denotes a $L\times 1$ vector specifying the location and $w(\s_i)$ is the spatial effect at location $\s_i$ with $\w = \w(\s) = (w(\s_1),w(\s_2),\ldots,w(\s_L))\trans$ denoting the $L\times 1$ vector of spatial effects; $\z_{ij}(\s_i)$ denotes a $q\times 1$ vector of known covariates for the dispersion model and $\bgamma$ is the corresponding $q\times 1$ vector of coefficients. A Bayesian hierarchical double generalized linear model (DGLM) using a non-canonical logarithmic link function is specified as
    \begin{equation}\label{eq:dglm_mean-disp}
         \log \mu_{ij}(\s_i) = \x_{ij}\trans(\s_i)\bbeta + \f_{ij}(\s_i)\trans \w(\s_i),~
            \log \phi_{ij} = \z_{ij}\trans\bgamma,
    \end{equation}
    which implies $\mu_{ij}(\s_i)=\mu_{ij}(\bbeta,\w)=\exp\left(\x_{ij}\trans(\s_i)\bbeta + \f_{ij}\trans(\s_i) \w(\s_i)\right)$ and $\phi_{ij}=\phi_{ij}(\bgamma) = \exp(\z_{ij}\trans\bgamma)$.
    
    \subsection{Hierarchical Prior Specification}\label{subsec:hps}
    In this section, we first present the hierarchical prior formulation for the model and process parameters, $\tilde{\btheta}$, followed by the prior formulation for the variable selection parameters $\btheta_{vs}$. Prior specification for model and process parameters are as follows:
    \begin{equation}\label{eq:jprior}
        \begin{split}
            &\text{Model Parameters: } \xi \sim U(a_{\xi},b_{\xi});\bbeta \sim N_p\left(\0_p,\blambda_\beta\trans\I_p\right); \bgamma \sim N_q\left(\0_q,\blambda_\gamma\trans\I_q\right),\\
            &\text{Process Parameters: } \phi_s \sim U\left(a_{\phi_s}, b_{\phi_s}\right);\sigma^{-2} \sim {\rm Gamma}\left(a_{\sigma}, b_{\sigma}\right);~ \nu\sim U(a_{\nu},b_{\nu}),\\
            &\text{Process: }\w \sim N_L\left(\0_L,\sigma^2\R(\phi_s)\right),
        \end{split}
    \end{equation}
	where $\0_m$ is the $m\times 1$ zero vector and $\I_m$ is the $m\times m$ identity matrix; $\X \in \mathbb{R}^{n \times p}$ and $\Z \in \mathbb{R}^{n \times q}$ are design matrices corresponding to the mean and dispersion models, with coefficients $\bbeta \in \mathbb{R}^{p}$ and $\bgamma \in \mathbb{R}^{q}$, respectively; $\F \in \mathbb{R}^{n \times L}$ is the spatial incidence matrix, $\w \in \mathbb{R}^{L}$ is the spatial effect, and $\R(\phi_s)=\sigma^2(\phi||\Delta||)^\nu K_\nu(\phi||\Delta||)$, where $K_\nu$ is the modified Bessel function of order $\nu$ \citep{abramowitz1988handbook}, is the Mat\'ern covariance kernel. Here $\{||\Delta||\}_{ii'}=||\s_i-\s_{i'}||_2$, the Euclidean distance between locations $\s_i$ and $\s_{i'}$. $U(\cdot\mid a,b)$ denotes the uniform distribution, $N_m(\cdot\mid\0,\Sigma)$ is the $m$-dimensional Gaussian with zero mean and covariance matrix $\Sigma$, and ${\rm Gamma}(\cdot\mid a,b)$ is the Gamma distribution with shape-rate parameterization. Note that the priors on $\blambda_{\beta} = (\lambda_{\beta,1},\ldots,\lambda_{\beta,p})$ and $\blambda_{\gamma} = (\lambda_{\gamma,1},\ldots,\lambda_{\gamma,q})$ are part of the variable selection priors and are discussed next. Referring to the framework in (\ref{eq:modelform}), the resulting posterior from (\ref{eq:jprior}) establishes the $[\text{data}\mid\text{process}, \btheta]\times [\text{process}\mid\widetilde{\btheta}]$ step. Conditional posteriors for $\widetilde{\btheta}$ are outlined in Section \ref{sec:post} of Supplementary Materials.
	
	For the continuous spike-and-slab prior formulation, $\btheta_{vs} = \{\zeta_\beta,\zeta_\gamma,\sigma^2_\beta,\sigma^2_\gamma,\alpha_\beta,\alpha_\gamma\}$ \citep[see, for e.g.,][]{ishwaran2005spike}. Note that we have separate prior formulations for mean and dispersion models. Let $\bbeta=(\beta_1,\beta_2,\ldots,\beta_p)\trans$ and $\bgamma=(\gamma_1,\gamma_2,\ldots,\gamma_q)\trans$ be the model coefficients corresponding to the mean and the dispersion models. Let us define $\lambda_{\beta,u} = \zeta_{\beta,u}\sigma_{\beta,u}^2$ and $\lambda_{\gamma,v} = \zeta_{\gamma,v}\sigma_{\gamma,v}^2$ for $u = 1,\ldots,p$ and $v = 1,\ldots,q$, respectively. We consider the following prior formulation:
	\begin{equation}
	\begin{aligned}\label{eq:spike-slab-c}
	    \pi(\btheta_{vs}) = \begin{cases} 
	    \zeta_{\beta,u}
     \stackrel{iid}{\sim} (1-\alpha_{\beta})\delta_{\nu_0}(\cdot)+\alpha_{\beta}\delta_{1}(\cdot),~
     \zeta_{\gamma,v}
     \stackrel{iid}{\sim} (1-\alpha_{\gamma})\delta_{\nu_0}(\cdot)+\alpha_{\gamma}\delta_{1}(\cdot),\\
	    \alpha_{\beta} \sim U(0,1),~\alpha_{\gamma} \sim U(0,1),\\
	    \sigma_{\beta,u}^{-2}\stackrel{iid}{\sim}{\rm Gamma}(a_{\sigma_\beta},b_{\sigma_\beta}),~\sigma_{\gamma,v}^{-2}\stackrel{iid}{\sim}{\rm Gamma}(a_{\sigma_\gamma},b_{\sigma_\gamma}),
	   \end{cases}
	\end{aligned}
	\end{equation}
	where $\beta_u$ and $\gamma_v$ have normal priors with mean 0 and variance $\zeta_{\beta,u}\sigma_{\beta,u}^2$ and $\zeta_{\gamma,u}\sigma_{\gamma,u}^2$, respectively. Here, $\delta_{c}(\cdot)$ denotes the discrete measure at $c$; hence, $\zeta_{\beta,u}$ and $\zeta_{\gamma,u}$ are indicators taking values $1$ or $v_0$ (small number close to 0) based on the selection of their corresponding covariates. The probabilities of these indicators taking the value 1 is given by $\alpha_{\beta}$ and $\alpha_{\gamma}$ respectively. We place a uniform prior on these selection probabilities and an inverse-Gamma prior on the parameters $\sigma_{\beta,u}^2$ and $\sigma_{\gamma,u}^2$. The choice of the shape and rate parameters ($a_{\sigma_\beta},b_{\sigma_\beta};a_{\sigma_\gamma},b_{\sigma_\gamma}$) of these inverse-Gamma priors induces a continuous bimodal distributions on $\zeta_{\beta,u}\sigma_{\beta,u}^2$ and $\zeta_{\gamma,u}\sigma_{\gamma,u}^2$ with a spike at $\nu_0$ and a right continuous tail. Combining the priors in (\ref{eq:jprior}) and (\ref{eq:spike-slab-c}) completes the hierarchical prior formulation for parameters $\btheta$ as defined in (\ref{eq:modelform}). Evidently, the above prior formulation allows for sufficient flexibility regarding variations in implementation. For instance, a hierarchical Bayesian framework for a simple DGLM can be obtained by omitting the process specification and variable selection. Analogously, DGLMs featuring variable selection or, spatial effects are obtained by omitting respective components from the prior specification outlined previously.
    \setlength{\tabcolsep}{10pt}
    \begin{table}[t]
	    \centering
	    \caption{Proposed Bayesian Hierarchical Double Generalized Linear Modeling Frameworks}
	    \label{tab:contrib}
	    \resizebox{\linewidth}{!}{
	        \begin{tabular}{l|l|c|c}
	    \hline
	    \hline
	       \multirow{2}{*}{Models} & \multirow{2}{*}{Frameworks}  &  \multirow{2}{*}{Specification ($\btheta$)} & \multirow{2}{*}{Number of Parameters}\\
	       &&\\\hline
	       \multirow{2}{*}{M1} & \multirow{2}{*}{DGLM} & \multirow{2}{*}{$\btheta_{m}$}&\multirow{2}{*}{$p+q+1$}\\
	       &&\\\hline
	       \multirow{2}{*}{M2} &\multirow{2}{*}{DGLM + Variable Selection} & \multirow{2}{*}{$\btheta_{m}$, $\btheta_{vs}$}&\multirow{2}{*}{$3p+3q+1$}\\
	       &&\\\hline
	       \multirow{2}{*}{M3} &\multirow{2}{*}{DGLM + Spatial Effect} & \multirow{2}{*}{$\btheta_{m}$, $\btheta_{pr}$}&\multirow{2}{*}{$p+q+L+4$}\\
	       &&\\\hline
	       \multirow{2}{*}{M4} &\multirow{2}{*}{DGLM + Spatial Effect + Variable Selection} & \multirow{2}{*}{$\btheta_{m}$, $\btheta_{vs}$, $\btheta_{pr}$}&\multirow{2}{*}{$3p+3q+L+4$}\\
	       &&\\
	    \hline
	    \hline
	    \end{tabular}
	    }
	\end{table}
 \subsection{Bayesian Estimation and Inference}
	In its full capacity (a model with spatial effects and variable selection) the model structure with prior specifications in (\ref{eq:jprior}) and (\ref{eq:spike-slab-c}) contains $3p+L+3q+4$ parameters. Depending on the dimensions of $\X(\s)$ and $\Z$, and the number of locations $L$, posterior inference can be a sufficiently daunting task. Traditional Metropolis--Hasting (M-H) random walk strategies are sub-optimal, involving costly pilot runs to determine viable initial starting points and unreasonably long chains while performing MCMC sampling. To avoid such issues, we use an adaptive rejection sampling while leveraging the log-concavity of the posteriors to perform effective inference that is not plagued by the above described issues \citep[for more details, see, for e.g,][]{girolami2011riemann}. In the following, we describe (i) briefly, our adaptive rejection MCMC sampling approach (more details are provided in the Supplementary Materials), (ii) the identifiability issues on the overall intercept that arise due to inclusion of a spatial effect and a strategy to address this, and (iii) a false discovery rate (FDR)--based approach for performing variable selection. 
	
	The joint posterior $\pi(\btheta\mid\y)$ generated as a result of the hierarchical priors in eq. \ref{eq:jprior} is sampled using a hybrid sampling strategy that includes M-H random walk and the Metropolis-Adjusted Langevin Algorithm (MALA) \citep{roberts2002langevin, girolami2011riemann}. We consider MALA updates for the model parameters $\{\bbeta,\w\}$ for the mean model. The dispersion model coefficients are sampled depending on the choice of likelihood, i.e., $\bgamma$ is sampled using a MALA if a saddle-point approximation of the likelihood is considered, otherwise $\bgamma$ is sampled using a MALA with a numerical approximation to the derivative of the conditional posterior for $\bgamma$ or using a M-H random walk. The parameters $\{\xi, \phi_s\}$ are updated using a M-H random walk. All the other remaining parameters are sampled using Gibbs sampling. In particular, we employ block updates for $\bbeta_{\w}=\{\bbeta,\w\}$ and $\bgamma$. Proposal variances feature adaptive scaling such that the optimal acceptance rate ($\approx 58\%$) to capture Langevin dynamics is achieved upon convergence \citep[see,][]{carlin2008bayesian,girolami2011riemann}. Proposal variances in the M-H updates also feature adaptive scaling such that the optimal acceptance rate ($\approx33\%$) for random walks is achieved upon convergence. We outline the full sampling algorithm at the end of Section \ref{sec:post} of the Supplement. For the hierarchical DGLM in (\ref{eq:dglm_mean-disp}), the specification of a spatial effect translates to fitting a random intercept mean model. Consequently, having an additional overall intercept $\beta_0$ in the model renders it unidentifiable \citep[see,][]{gelfand1995efficient,gelfand1996efficient}. Hence, $\beta_0$ is not estimable, although $\beta_0+\w$ is estimable. $\beta_0$ is estimated 
    through hierarchical centering of the posterior for $\w$ \citep[see, for e.g.,][]{gelfand1996efficient}.
	
	The MCMC samples of $\bbeta$ and $\bgamma$ explore their conditional posterior distributions and point estimates for these model parameters can be obtained using maximum a-posteriori (MAP) estimates or the posterior means. Although we obtain point estimates, these estimates do not yield exact zero values since we have considered a continuous spike-and-slab prior with a spike at $\nu_0$ (a small positive number). Additionally, these estimates do not make use of the all the MCMC samples. We use a Bayesian model averaging--based strategy that leverages all the MCMC samples to build inference \citep{hoeting1999bayesian}. Specifically, we use a FDR-based using Bayesian model averaging combined with point estimates \citep[see for e.g.,][]{morris2008bayesian,mohammed2021radiohead}. Let $\beta_u^{(m)}$ for $m=1,\ldots,M$ denote the MCMC samples (after burn-in and thinning) for the coefficients of the mean model. We compute $p_u = \frac{1}{M}\sum_m I\big(|\beta_u^{(m)}| \leq c\big)$, where $I(\cdot)$ is the indicator function; these probabilities $p_u$ can be interpreted as local FDR \citep{morris2008bayesian}. The probability $(1-p_u)$ can be interpreted as the probability that covariate $u$ is significantly associated with the response. We use $p_u$s to decide on which covariates to select while controlling the FDR at level $\alpha$. That is, we infer that the covariate $u$ has a {\em non-zero} coefficient if $p_u < \kappa_\alpha$ for some threshold $\kappa_\alpha \in (0,1)$. We compute the threshold $\kappa_\alpha$ as follows: We first sort the probabilities $p_u$ and denote the sorted probabilities as $p_{(u)}$ for $u=1,\ldots,p$. We then assign $\kappa_\alpha = p_{(u^*)}$, where $u^* = \max\left\{\widetilde{u} \mid \frac{1}{\widetilde{u}}\sum_{u=1}^{\widetilde{u}} p_{(u)} \leq \alpha\right\}$. This approach caps our false discoveries of selected variables at $100\alpha\%$. We employ the same approach using the MCMC samples of $\bgamma$ to select significant coefficients for the dispersion models.

 Posterior inference on $\bbeta$, $\bgamma$ is performed using MAP (point) estimates along with posterior mean, median, standard deviation and highest posterior density (HPD) intervals. For $\w$, we employ posterior mean, median, standard deviation and HPD intervals to perform inference. Next, we demonstrate some synthetic experiments that document the performance of our proposed models. The computation has been performed in the {\tt R} statistical environment. The required subroutines can be accessed via an open-source repository at: \url{https://github.com/arh926/sptwdglm}.
	\begin{table}[t]
	\caption{Parameter settings used to obtain varying proportion of zeros in the synthetic data.}\label{tab:synexp-1}
    \centering
    \begin{tabular}{c|@{\extracolsep{20pt}}ccc@{}}
    \hline
    \hline
    \multirow{2}{*}{Proportion of 0s} & \multirow{2}{*}{$\mu_{\bbeta_{-0}} \left(\sigma_{\bbeta_{-0}}\right)$} & \multirow{2}{*}{$\gamma_0$} & \multirow{2}{*}{$\mu_{\bgamma_{-0}} \left(\sigma_{\bgamma_{-0}}\right)$} \\ 
    &&&\\
    \hline
    15\% & 0.50 (0.1) & -1.50 & 0.50 (0.1) \\ 
    30\% & 0.50 (0.1) & 0.70 & 0.50 (0.1) \\ 
    60\% & 0.50 (0.1) & 2.50 & 0.50 (0.1) \\ 
    80\% & 1.00 (0.1) & 4.50 & 0.50 (0.1) \\ 
    95\% & 1.00 (0.1) & 7.00 & 0.50 (0.1) \\ 
    \hline
    \hline
    \end{tabular}
    \end{table}
    \section{Synthetic Experiments}\label{sec:synexp}
\begin{figure}[htb]
    \centering
    \includegraphics[width=0.8\linewidth, height=0.6\linewidth]{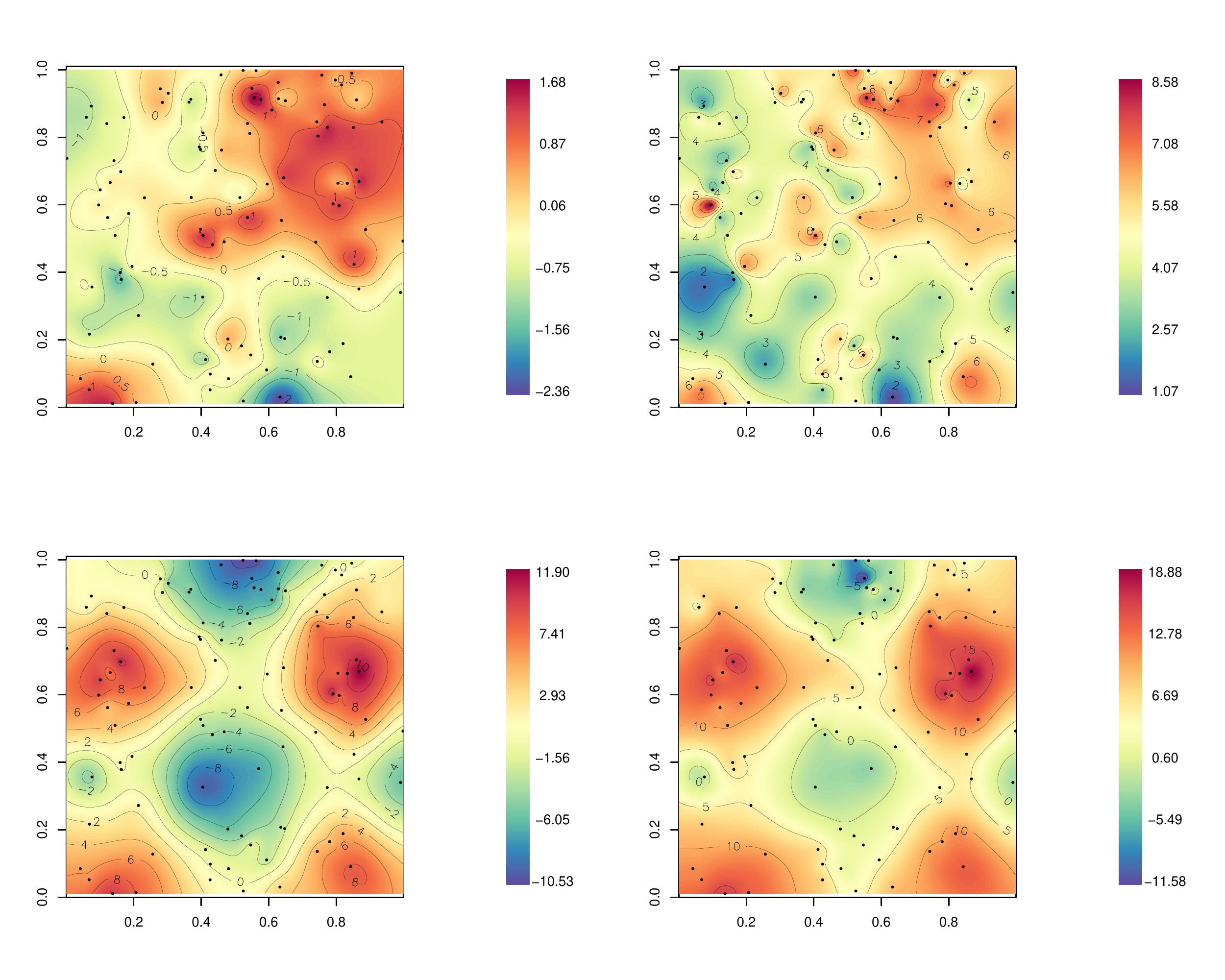}
    \caption{\small Plots showing synthetic spatial patterns, pattern 1 (top, left column) and pattern 2 (bottom, left column) and corresponding logarithm of aggregated synthetic response (right column).}
	\label{fig::synexp}
    \end{figure}

    We begin with an observation---the spatial heterogeneity that our models aim to quantify is not observed in real life. Hence, it is imperative to document the accuracy of estimating such effects through synthetic experiments. Settings used are outlined---we consider varying proportion of zeros (15\%, 30\%, 60\%, 80\% and 95\%) under which the quality of posterior inference for $\btheta$ is assessed. Proportion of zeros can be interpreted as an inverse signal-to-noise ratio for the synthetic response. For the sake of brevity, we only show the results for synthetic experiments pertaining to Bayesian variable selection in the presence of spatial effects. Additional simulations can be found in the online Supplement. 
    To construct the synthetic data we consider three scenarios pertaining to {\em model structure}, (a) there is no overlap (i.e. selected $\beta$'s and $\gamma$'s do not intersect) (b) 50\% overlap (in the union of all selected variables across the mean and dispersion models) (c) 100\% overlap between mean and dispersion model specification. We use 10 covariates including an intercept, where the columns of the synthetic design matrices $\X$ and $\Z$ are hierarchically centered and scaled, independently sampled Gaussian variables with mean 0 and variance 1. Naturally, specification of the true $\bbeta$, $\w$ and $\bgamma$ parameters determine the proportion of zeros in the synthetic response. Table \ref{tab:synexp-1} contains the parameter specifications used. The true value of the index parameter, $\xi = 1.5$. In an attempt to produce a synthetic setup that resembles reality we simulate, $\w\sim N(5(\sin(3\pi\s_{1})+\cos(3\pi\s_2)),1)$ (see Figure \ref{fig::synexp}, second row). The alternative route would be to fix values of $\sigma^2$ and $\phi_s$ and generate a realization $\w \sim N_{L}(\0_L,\sigma^2\R(\phi_s))$ (see Figure \ref{fig::synexp}, first row). Under each setting we consider $M = 10$ replications. Within each replication we fit all the proposed modeling frameworks as shown in Table \ref{tab:contrib}. 
    The hyper-parameter settings used while specifying priors for the models are, $a_{\xi} = 1$, $b_{\xi} = 2$, $a_\sigma = a_{\sigma_{\beta}} = a_{\sigma_{\gamma}} = 2$, $b_\sigma = b_{\sigma_{\beta}} = b_{\sigma_{\gamma}}=1$ (producing inverse-Gamma priors with mean 1 and infinite variance), $a_{\phi_s} = 0$, $b_{\phi_s}=30$, $\sigma^{-2}_{\beta} = \sigma^{-2}_{\gamma} = 10^{-6}$ and $\nu_0=5\times 10^{-4}$, producing a vague and non-informative hierarchical prior. We maintain an FDR of 5\% for all settings while performing model selection. The sample size varies from $N=\{2\times 10^3, 5\times 10^3, 1\times 10^4\}$ and the number of locations are $L=1\times10^2$. Across replications, false positive rate (FPR) and true positive rate (TPR) are computed to measure accuracy of our model selection procedure. To record the quality of estimation, we compute the mean squared error (MSE), for e.g. $MSE(\bbeta) = \frac{1}{p}\sum_{i_\beta = 1}^{p}(\beta_{i_\beta}-\widehat{\beta}_{i_\beta})^2$, which can be computed similarly for the other parameters. We also compute average coverage probabilities, for e.g. considering these probabilities for $\bbeta$ we define $CP(\bbeta) =\frac{1}{M} \sum_{m=1}^{M}I(\bbeta_{true}\in (l_m(\bbeta),u_m(\bbeta)))$, where $l_m(\bbeta)$ and $u_m(\bbeta)$ are the lower and upper 95\% HPD intervals respectively for $\bbeta$ in replication $m$; we obtain coverage probabilities for $\w$ and $\bgamma$ similarly. The results obtained under the above settings are shown in Table \ref{tab:res-sim}. The first column is named configuration (abbreviated as config.) with entries denoting the proportion of overlap between selected coefficients in the mean and dispersion models, which is indicative of model structure. This is estimated by observing the overlap between selected variables following the model fit (for models M2 and M4). No variable selection is performed for models M1 and M3. 

    \begin{figure}[t]
	\centering
	\includegraphics[width=1\linewidth]{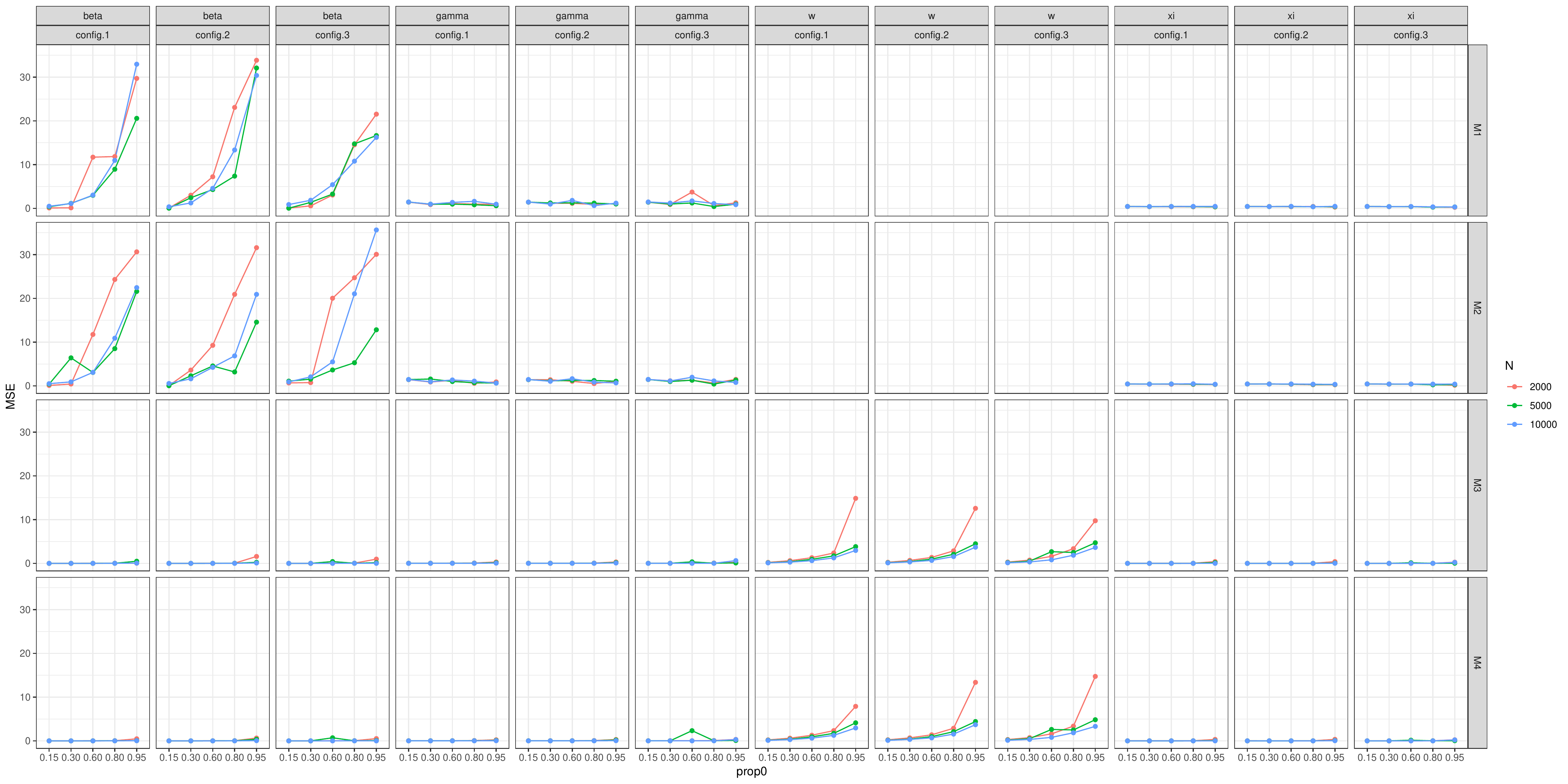}
	\caption{\small Results for synthetic experiments showing model performance, where MSE is plotted against proportion of zeros in the synthetic response which is tabulated across models (M1, M2, M3, M4) vs. $\{\btheta_m,\btheta_{pr}\}\times$ configuration.}
	\label{fig::res-synexp}
    \end{figure}
    
    \begin{table}[t]
    \centering
    \caption{\small Table showing results of synthetic experiments for model selection for models M2 and M4. Corresponding standard deviations are shown in brackets below.}\label{tab:res-sim}
    \resizebox{\linewidth}{!}{
    \begin{tabular}{c|c|c|ccc|ccc}
      \hline
      \hline
      \multirow{2}{*}{$N$} & \multirow{2}{*}{True Overlap} & \multirow{2}{*}{Prop. of 0s} & \multicolumn{3}{c|}{M2} & \multicolumn{3}{c}{M4}\\
      \cline{4-6}\cline{7-9}
      & & & Overlap & FPR & TPR & Overlap & FPR & TPR \\
      \hline
      \multirow{24}{*}{5000} &\multirow{8}{*}{0.00} & \multirow{2}{*}{0.15} & 0.03 & 0.04 & 0.69 & 0.00 & 0.00 & 0.90 \\
      & & & (0.06) & (0.06) & (0.07) & (0.00) & (0.00) & (0.00) \\ \cline{3-9}
      & & \multirow{2}{*}{0.30} & 0.05 & 0.05 & 0.89 & 0.00 & 0.00 & 1.00 \\ 
      & & & (0.09) & (0.09) & (0.10) & (0.00) & (0.00) & (0.00) \\ \cline{3-9}
      & & \multirow{2}{*}{0.60} & 0.01 & 0.01 & 0.98 & 0.00 & 0.00 & 1.00 \\ 
      & & & (0.04) & (0.04) & (0.04) & (0.00) & (0.00) & (0.00) \\ \cline{3-9}
      & & \multirow{2}{*}{0.80} & 0.04 & 0.04 & 0.99 & 0.00 & 0.00 & 1.00 \\ 
      & & & (0.06) & (0.06) & (0.03) & (0.00) & (0.00) & (0.00) \\ \cline{3-9}
      & & \multirow{2}{*}{0.95} & 0.10 & 0.06 & 0.89 & 0.08 & 0.04 & 0.95 \\ 
      & & & (0.20) & (0.10) & (0.20) & (0.10) & (0.11) & (0.08) \\ \cline{2-9}
      & \multirow{8}{*}{0.50} & \multirow{2}{*}{0.15} & 0.24 & 0.01 & 0.67 & 0.50 & 0.00 & 0.90 \\ 
      & & & (0.16) & (0.05) & (0.09) & (0.00) & (0.00) & (0.03) \\ \cline{3-9}
      & & \multirow{2}{*}{0.30} & 0.49 & 0.01 & 0.94 & 0.50 & 0.00 & 1.00 \\ 
      & & & (0.08) & (0.05) & (0.06) & (0.00) & (0.00) & (0.00) \\\cline{3-9}
      & & \multirow{2}{*}{0.60} & 0.51 & 0.00 & 0.93 & 0.50 & 0.00 & 1.00 \\ 
      & & & (0.03) & (0.00) & (0.06) & (0.00) & (0.00) & (0.00) \\ \cline{3-9}
      & & \multirow{2}{*}{0.80} & 0.56 & 0.07 & 0.97 & 0.49 & 0.01 & 1.00 \\ 
      & & & (0.09) & (0.08) & (0.04) & (0.02) & (0.05) & (0.00) \\ \cline{3-9}
      & & \multirow{2}{*}{0.95} & 0.65 & 0.20 & 0.85 & 0.48 & 0.04 & 0.92 \\ 
      & & & (0.31) & (0.17) & (0.22) & (0.10) & (0.15) & (0.18) \\ \cline{2-9}
      & \multirow{8}{*}{1.00} & \multirow{2}{*}{0.15} & 0.40 & 0.03 & 0.66 & 1.00 & 0.00 & 0.90 \\ 
      & & & (0.25) & (0.05) & (0.07) & (0.00) & (0.00) & (0.00) \\ \cline{3-9}
      & & \multirow{2}{*}{0.30} & 0.90 & 0.04 & 0.86 & 1.00 & 0.00 & 1.00 \\ 
      & & & (0.14) & (0.08) & (0.08) & (0.00) & (0.00) & (0.00) \\ \cline{3-9}
      & & \multirow{2}{*}{0.60} & 1.00 & 0.00 & 0.96 & 1.00 & 0.00 & 1.00 \\ 
      & & & (0.00) & (0.00) & (0.05) & (0.00) & (0.00) & (0.00) \\ \cline{3-9}
      & & \multirow{2}{*}{0.80} & 1.00 & 0.00 & 1.00 & 1.00 & 0.00 & 1.00 \\ 
      & & & (0.00) & (0.00) & (0.00) & (0.00) & (0.00) & (0.00) \\ \cline{3-9}
      & & \multirow{2}{*}{0.95} & 0.35 & 0.15 & 0.85 & 0.89 & 0.10 & 0.89 \\ 
      & & & (0.21) & (0.16) & (0.22) & (0.21) & (0.11) & (0.10) \\
   \hline
   \hline
    \end{tabular}
    }
    \end{table}
From the results shown, we see that models M1 and M2 perform poorly. Estimates, $\widehat{\bbeta}$ remain fairly unaffected as compared to $\widehat{\bgamma}$ and $\widehat{\xi}$, where all of the variation not quantified, yet present in the synthetic data spills over to corrupt and compromise the quality of estimates. This also does not produce reliable results pertaining to model structure recovery for M1 and M2. However, significant improvements show up with M3 and M4. Particularly, under higher proportion of zeros in the synthetic data (low signal to noise ratio) the performance of M4 remains stable with respect to model structure recovery and estimation of parameters (refer to Table \ref{tab:res-sim}), thereby producing {\em robust inference} among the models in comparison. As an example within our simulation setting, under 95\% of 0s in the data and under low sample sizes, for example 2000 or, 5000, the estimates of model coefficients and spatial effects in M3 and M4 are adversely affected by locations having fewer non-zero observations. This observation addresses the concern around specifying DGLMs without spatial random effects in a scenario where the data displays spatial variation. The results demonstrate expected gains when our model in its full capacity is used instead of an usual DGLM.
	
	We use the MCMC algorithm featuring MALA updates for $\bbeta, \w$  and $\bgamma$. Chain lengths are set to $1\times10^4$, with the initial 5,000 samples as burn-in and thin the rest by selecting every 10-th sample which reduces any remaining auto-correlation and produces 500 independent posterior samples for each setting. The posterior estimate, $\widehat{\btheta}$ is then obtained using the produced samples by computing the median or a MAP estimate as applicable for the model.  Coverage probabilities for model M4 remained sufficiently high ($\approx 1$) across all settings; only declining marginally for $\w$ (remaining above $90\%$) under high proportions of zeros (low signal to noise ratio) in the data. We performed additional synthetic experiments to showcase (a) the performance of M3 with respect to the quality of estimation for spatial effects and (b) the performance of M2. They are detailed in the Supplementary materials---we briefly outline its contents in the next section.
 
    \section{Supplementary Analysis}\label{sec:supp}
    The online Supplement to this paper contains details on the derivations of the posteriors essential for constructing MCMC subroutines. They are outlined in Section \ref{sec:post}. Section \ref{sec:ede} features additional simulation experiments that supplement those outlined previously in Section \ref{sec:synexp}. It documents performance of M2, shown in Table \ref{tab:synexp-11}, contains results of experiments for scenarios featuring spatial covariates, shown in Table \ref{tab:sim-spcov}, and varying spatial patterns as seen in Figure \ref{fig::synexp}, shown in Tables \ref{tab:sim-pat-1} and \ref{tab:sim-pat-2}. Convergence diagnostics are shown for selected model parameters (index parameter $\xi$) in Section \ref{subsec:conv}. Contents of the \texttt{R}-package are described in Section \ref{subsec:r-pack}. Finally, results for models M1 and M3 pertaining to the real data analysis described in the next section appear in Section \ref{sec:supp-real}, Tables \ref{tab:mc-mean-dglm-m1}, \ref{tab:mc-disp-dglm-m1}, \ref{tab:mc-mean-spdglm-m3} and \ref{tab:mc-disp-spdglm-m3}.

    \section{Automobile Insurance Premiums, Connecticut, 2008}\label{sec:app}
	
	\begin{figure}[t]
	\centering
	\includegraphics[width=0.95\linewidth, height = 0.4\linewidth]{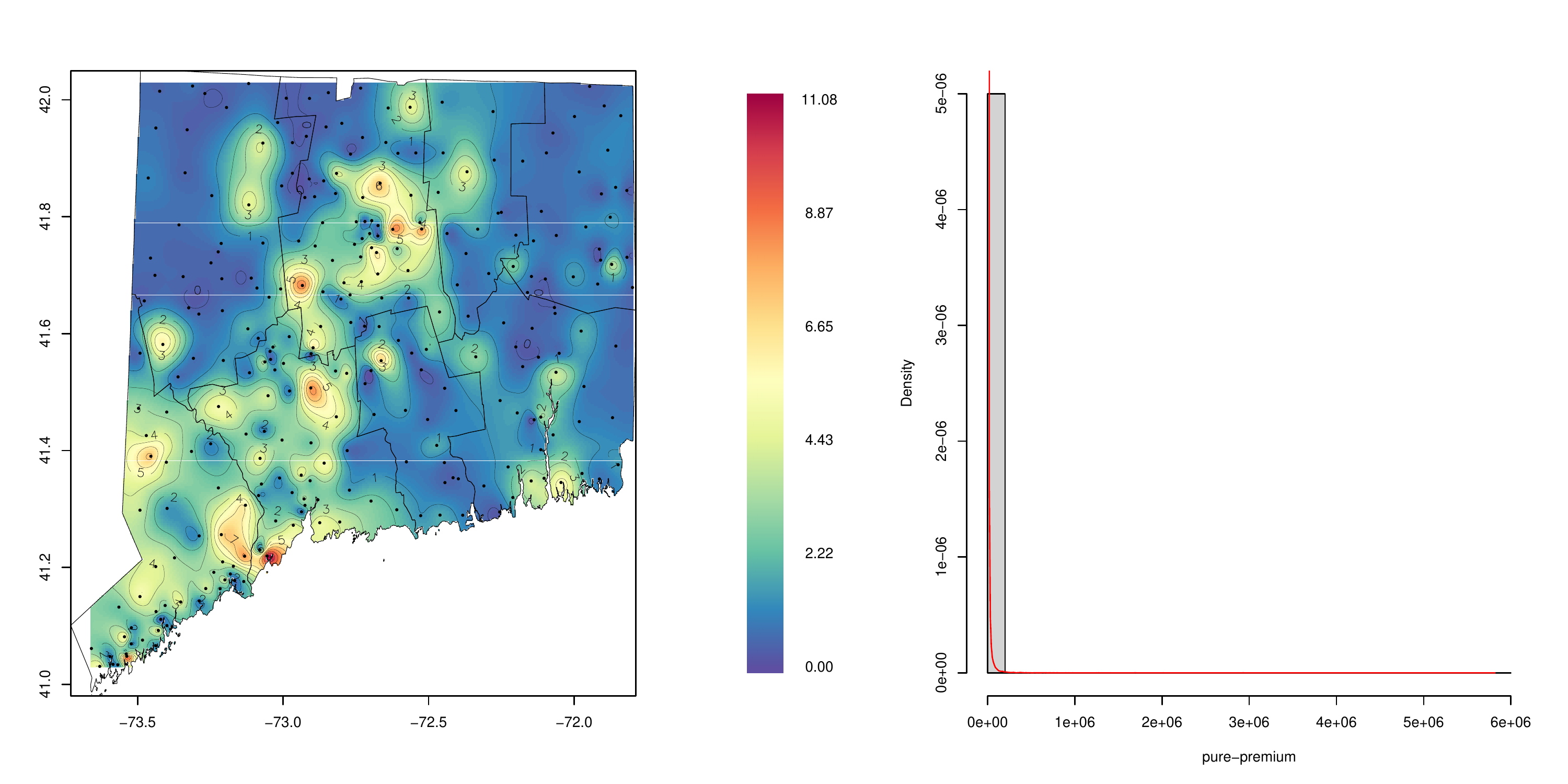}
	\caption{\small (left) Spatial plot of zip-code level aggregated pure-premium $\times 10^{-6}$ for the state of Connecticut, 2008 (right) histogram for the pure-premium overlaid with a probability density estimate.}
	\label{fig::ct-2008}
    \end{figure}
    
	We analyze automobile insurance premiums for collision coverage in the state of Connecticut, for the year 2008. The data is obtained from a comprehensive repository named Highway Loss Data Institute (HLDI) maintained by the independent non-profit, Insurance Institute for Highway Safety (IIHS) (\url{https://www.iihs.org/}) working towards reducing losses arising from motor vehicle collisions in North America. We briefly describe the variables contained in HLDI data. It records covariate information at three levels for an insured individual. They are as follows,
 \begin{itemize}
    \item [a. ] {\it Individual Level:} (i) accident and model year of the vehicle, (ii) age, gender, marital status.
    \item [b. ] {\it Policy level:} (i) policy payments, measured in United States dollars, (ii) exposure--measured in policy years, for e.g. 0.5 indicates a coverage period of 6 months or, half a year, measured in years, (iii) policy risk--having two levels, which is assigned by the insurance company based on provided information by the individual, (iv) deductible limit--with 8 categories.
    \item[c. ] {\it Spatial:} 5-digit zip code.
    \end{itemize} 
    Derived variables like age categories, vehicle age in years and interactions like gender $\times$ marital status are computed and used as covariates in the model. For the state of Connecticut, 1,513,655 ($\approx$ 1.5 million) data records were obtained in the year 2008, at 281 zip-codes. Zip-codes are areal units, we consider the latitude-longitude corresponding to the centroid of each zip code as the point reference counterpart unit for our application purposes. Distance between two zip-codes is then specified as the Euclidean distance between their centroids. The proportion of zeros in the payments is 95.73\%. From an insurer's perspective, policy rate-making is the problem of assigning policy-premium to a new customer's policy based on their covariate information (for instance, individual level, policy and residence zip-code). We achieve this via out-sample prediction. To that end, we consider a 60-40 split for the data, the split is performed by using stratified sampling without replacement over zip-codes, such that the same 281 zip-codes are also available in the out-sample. The training data then contains $N_{tr} = 908,741$ observations with $N_{pr} = 604,914$ observations kept in reserve for prediction constituting the out-sample data.

    We denote payments towards a policy made by individual $i$, residing in zip-code $j$ as $y_{ij}$ with an exposure of $t_{ij}$. We assume that the policy-premium defined as $y^*_{ij}= \frac{y_{ij}}{t_{ij}}\sim Tw\left(\mu_{ij}, \phi_{ij}, \xi\right)$, which implies $ y_{ij} \sim Tw(t_{ij}\mu_{ij}, t_{ij}^{2 - \xi}\phi_{ij}, \xi)$ using the scale invariance property. The following hierarchical DGLM is then specified,
    	\begin{align}\label{eq:spdglm-ct}
     \begin{split}
    	     \log \mu_{ij}(\s_i) &= -\log t_{ij} + \x_{ij}\trans(\s_i)\bbeta + \f_{ij}(\s_i)\trans \w(\s_i),\\
          \log \phi_{ij} &= -(2-\xi)\log t_{ij} + \z_{ij}\trans\bgamma,
     \end{split}
    	\end{align}
    when considered with a spatial specification, where the terms $-\log t_{ij}$ and $-(2-\xi)\log t_{ij}$ act as offsets for the respective mean and dispersion models. Given the covariates described at the beginning $p = q = 29$, producing a $N_{tr} \times (p-1)$ design matrix for the mean model and a $N_{tr} \times q$ design matrix for the dispersion model. The model in (\ref{eq:spdglm-ct}) specifies model M3 from Table \ref{tab:contrib}, M4 is obtained by specifying $\pi(\btheta_{vs})$ from (\ref{eq:spike-slab-c}) on $\btheta_m$. M1 is obtained by setting $f_{ij}(\s_i) = \0$ and M2 is obtained by specifying $\pi(\btheta_{vs})$ on $\btheta_m$ for the resulting model. For M1 and M2 we include an intercept in the mean model. We fit models M1--4 on the training data. Model selection is performed using FDR based variable selection on the posterior MCMC samples from fitting models M2 and M4, controlling for FDR at 1\%. The performance of M1--4 is assessed using the Akaike Information Criteria (AIC) \citep[][]{akaike1998information}.
    \begin{figure}[t]
	\begin{subfigure}[b]{0.45\linewidth}
	\centering
                \includegraphics[width=1\linewidth, height = 0.9\linewidth]{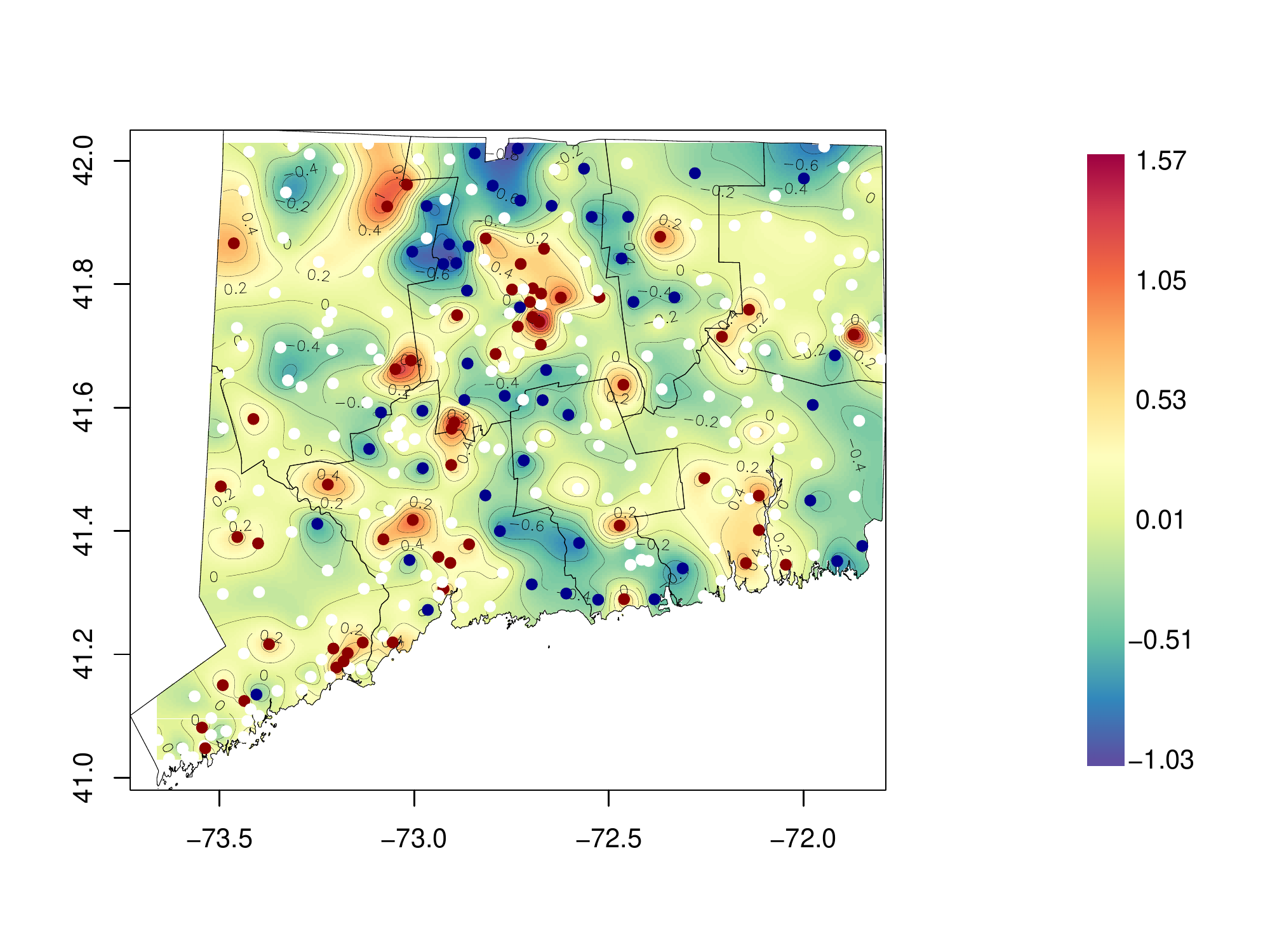}
    \end{subfigure}%
    \begin{subfigure}[b]{0.45\linewidth}
    \centering
                \includegraphics[width=1\linewidth, height = 0.9\linewidth]{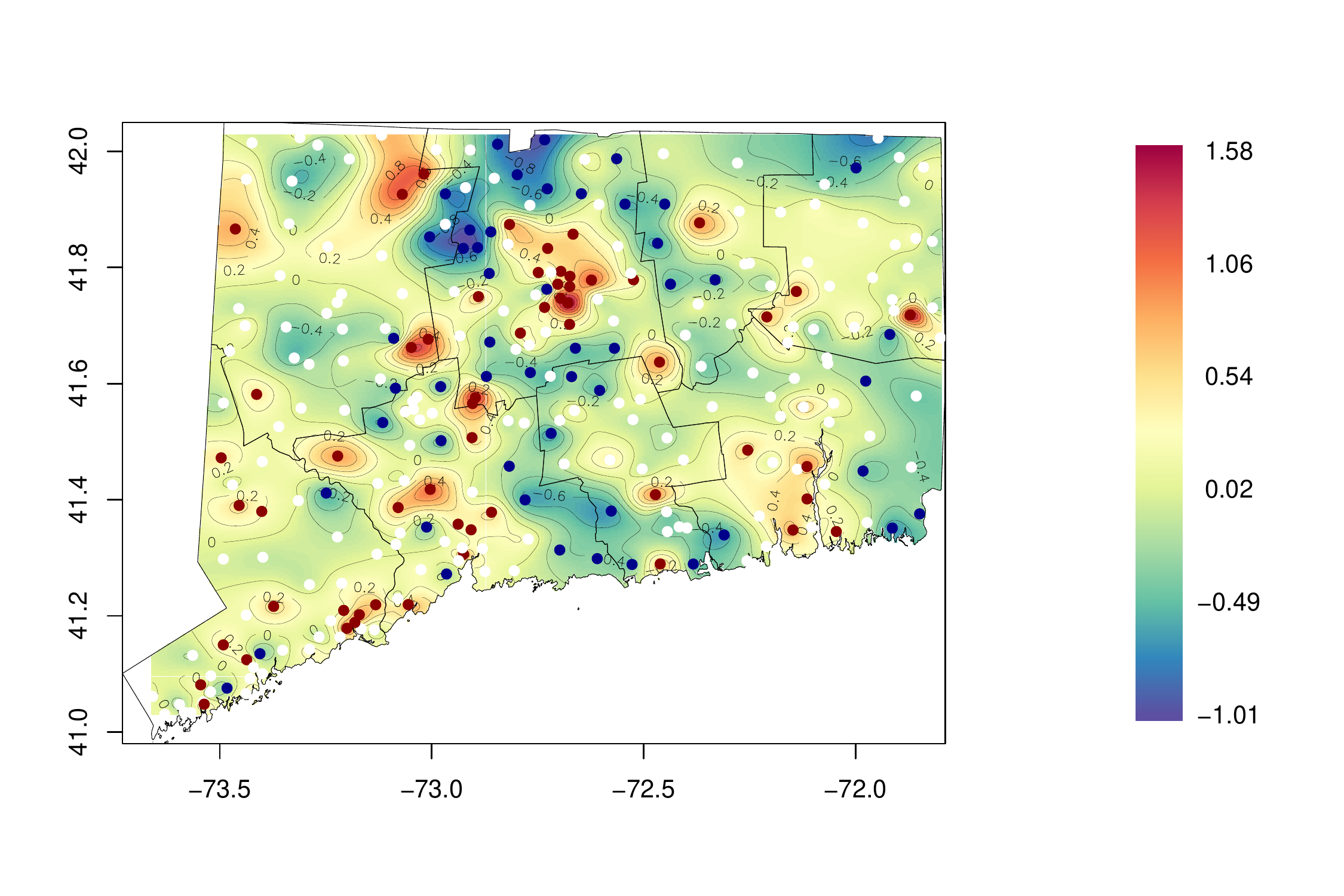}
    \end{subfigure}%
	\caption{\small Spatial plot showing the posterior estimates for 281 zipcodes from (left) M3, and (right) M4 in Connecticut. The locations are color coded based on significance, with white indicating a location with 0 in its HPD interval, blue (red) indicating HPD interval with both endpoints negative (positive).}
	\label{fig::ct-fit-08}
    \end{figure}
    While specifying (\ref{eq:jprior}) and (\ref{eq:spike-slab-c}) the hyper-parameter settings used are, $\sigma^2_{\beta}=\sigma^2_{\gamma}=10^6$, $a_{\sigma_\beta}=a_{\sigma}=a_{\sigma_\gamma}=2$ and $b_{\sigma_\beta}=b_{\sigma}=b_{\sigma_\gamma}=1$ (generating inverse-gamma priors with mean 1 and infinite variance), $a_{\phi_s}=0$, $b_{\phi_s}=60$. We maintain $a_{\xi}=1$ and $b_{\xi}=2$ for all models. For ease of implementation, the fractal parameter, $\nu$ is fixed at 0.5, producing the exponential covariance kernel. We consider $1\times 10^5$ MCMC iterations for generating samples from respective posteriors with burn-in diagnosed at $5\times 10^4$ and include every 20-th sample to compute posterior estimates as our thinning strategy. Convergence is assessed through inspecting trace plots. Proposal variances were scaled in an adaptive fashion to provide optimal acceptance rates of 58\% (MALA) and 33\% (MH). Predictive performance for models M1--4 is judged based on square root deviance on the out-sample data. The results are shown in Table \ref{tab:tab-res}. Optimal values are marked in bold. We show the results for estimated model coefficients featuring Bayesian variable selection (models M2 and M4) in Tables \ref{tab::mc-ssdglm} and \ref{tab::mc-spssdglm}. Results for M1 and M3 are postponed to Tables \ref{tab:mc-mean-dglm-m1}, \ref{tab:mc-disp-dglm-m1}, \ref{tab:mc-mean-spdglm-m3} and \ref{tab:mc-disp-spdglm-m3}
    in the Supplementary Materials. Posterior estimates for the spatial effects in models M3 and M4 are shown in Figure \ref{fig::ct-fit-08}. Zip-codes with significant effects are color coded appropriately. 
    \begin{table}[!]
    \centering
    \caption{Estimated Coefficients for fixed effects corresponding to model M2. We show the maximum a-posteriori (MAP) estimates along with median, mean, standard deviation and highest posterior density intervals (HPD). The variables listed are a result of FDR-based variable selection at 1\%.}\label{tab::mc-ssdglm} 
    \resizebox{\linewidth}{!}{
       \begin{tabular}{c|c|c|@{\extracolsep{4em}}cccccc@{}}
       \hline
       \hline
       \multicolumn{2}{c|}{Parameters} & Levels & MAP & Median & Mean & SD & Lower HPD & Upper HPD \\ 
       \hline
       \hline
      \multirow{15}{*}{$\bbeta$} & (Intercept) & -- & 5.815 & 5.965 & 5.955 & 0.151 & 5.706 & 6.227 \\ \cline{2-3}
      & \multirow{3}{*}{age.car} & 1 & 0.113 & 0.112 & 0.110 & 0.033 & 0.044 & 0.173 \\ 
      & & 2 & 0.234 & 0.239 & 0.240 & 0.030 & 0.186 & 0.304 \\ 
      & & 5 & 0.124 & 0.122 & 0.121 & 0.033 & 0.062 & 0.189 \\ \cline{2-3} 
      & \multirow{1}{*}{risk} & S & -0.213 & -0.217 & -0.217 & 0.023 & -0.258 & -0.168 \\ \cline{2-3}
      & \multirow{3}{*}{agec} & 3 & -0.399 & -0.399 & -0.400 & 0.030 & -0.454 & -0.341 \\ 
      & & 4 & -0.518 & -0.515 & -0.515 & 0.035 & -0.580 & -0.444 \\ 
      & & 5 & -0.712 & -0.710 & -0.708 & 0.065 & -0.832 & -0.575 \\ \cline{2-3} 
      & \multirow{2}{*}{gender} & F & 0.493 & 0.502 & 0.513 & 0.067 & 0.400 & 0.648 \\ 
      & & M & 0.608 & 0.614 & 0.619 & 0.053 & 0.521 & 0.731 \\ \cline{2-3} 
      & \multirow{1}{*}{marital} & M & -0.376 & -0.392 & -0.403 & 0.067 & -0.577 & -0.302 \\ \cline{2-3} 
      & \multirow{3}{*}{deductible} & B & 0.898 & 1.051 & 1.104 & 0.244 & 0.740 & 1.525 \\ 
      & & E & 0.616 & 0.482 & 0.478 & 0.158 & 0.192 & 0.741 \\ 
      & & F & 0.580 & 0.428 & 0.435 & 0.154 & 0.135 & 0.656 \\ 
      & & G & 0.287 & 0.348 & 0.358 & 0.159 & 0.073 & 0.633 \\ \hline
      \multirow{18}{*}{$\bgamma$} & (Intercept) & -- & 7.354 & 7.346 & 7.345 & 0.048 & 7.249 & 7.435 \\ \cline{2-3}
      & \multirow{8}{*}{age.car} & 0 & -0.820 & -0.811 & -0.811 & 0.053 & -0.911 & -0.714 \\ 
      & & 1 & -1.024 & -1.017 & -1.016 & 0.051 & -1.115 & -0.922 \\ 
      & & 2 & -0.888 & -0.874 & -0.873 & 0.052 & -0.969 & -0.776 \\ 
      & & 3 & -0.864 & -0.854 & -0.851 & 0.052 & -0.953 & -0.760 \\ 
      & & 4 & -0.843 & -0.838 & -0.838 & 0.052 & -0.936 & -0.743 \\ 
      & & 5 & -0.788 & -0.781 & -0.780 & 0.052 & -0.877 & -0.682 \\ 
      & & 6 & -0.770 & -0.765 & -0.765 & 0.053 & -0.862 & -0.665 \\ 
      & & 7 & -0.730 & -0.723 & -0.723 & 0.053 & -0.829 & -0.627 \\ \cline{2-3} 
      & \multirow{1}{*}{risk} & S & 0.114 & 0.114 & 0.114 & 0.011 & 0.093 & 0.135 \\ \cline{2-3} 
      & \multirow{1}{*}{agec} & 5 & -0.289 & -0.293 & -0.293 & 0.029 & -0.345 & -0.234 \\ \cline{2-3}
      & \multirow{6}{*}{deductible} & B & -0.971 & -0.963 & -0.951 & 0.099 & -1.136 & -0.759 \\ 
      & & C & -0.519 & -0.522 & -0.517 & 0.049 & -0.614 & -0.414 \\ 
      & & D & -0.565 & -0.568 & -0.564 & 0.044 & -0.657 & -0.470 \\ 
      & & E & -0.506 & -0.501 & -0.495 & 0.042 & -0.568 & -0.394 \\ 
      & & F & -0.348 & -0.346 & -0.340 & 0.039 & -0.410 & -0.242 \\ 
      & & H & 0.411 & 0.419 & 0.426 & 0.097 & 0.251 & 0.618 \\ \cline{2-3} 
      & \multirow{1}{*}{genderMarital} & A & -0.113 & -0.106 & -0.097 & 0.043 & -0.168 & -0.001 \\ \hline
      $\xi$ & -- & -- & 1.673 & 1.673 & 1.673 & 0.001 & 1.671 & 1.676 \\ 
       \hline
       \hline
    \end{tabular}
    }
    \end{table}

    \begin{table}[!]
    \centering
    \caption{Estimated coefficients for fixed effects corresponding to model M4. We show the MAP, median, mean, standard deviation and HPDs. The variables listed are a result of FDR-based variable selection at 1\%.}\label{tab::mc-spssdglm} 
    \resizebox{\linewidth}{!}{
        \begin{tabular}{c|c|c|@{\extracolsep{4em}}cccccc@{}}
      \hline
      \hline
      \multicolumn{2}{c|}{Parameters} & Levels & MAP & Median & Mean & SD & Lower HPD & Upper HPD \\ 
      \hline
      \multirow{23}{*}{$\bbeta$}  & (Intercept) & -- & 5.215 & 5.213 & 5.216 & 0.135 & 5.165 & 5.327 \\ \cline{2-3}
      & \multirow{8}{*}{age.car} & 0 & 0.252 & 0.251 & 0.251 & 0.063 & 0.123 & 0.377 \\ 
      & & 1 & 0.329 & 0.317 & 0.318 & 0.058 & 0.217 & 0.431 \\ 
      & & 2 & 0.424 & 0.423 & 0.425 & 0.061 & 0.318 & 0.549 \\ 
      & & 3 & 0.209 & 0.193 & 0.195 & 0.060 & 0.084 & 0.318 \\ 
      & & 4 & 0.236 & 0.252 & 0.255 & 0.062 & 0.149 & 0.387 \\ 
      & & 5 & 0.297 & 0.302 & 0.307 & 0.059 & 0.191 & 0.428 \\ 
      & & 6 & 0.182 & 0.209 & 0.214 & 0.061 & 0.113 & 0.343 \\ 
      & & 7 & 0.155 & 0.154 & 0.157 & 0.063 & 0.049 & 0.286 \\ \cline{2-3}
      & \multirow{1}{*}{risk} & S & -0.190 & -0.184 & -0.183 & 0.023 & -0.226 & -0.137 \\ \cline{2-3}
      & \multirow{4}{*}{agec} & 2 & -0.121 & -0.126 & -0.128 & 0.030 & -0.186 & -0.068 \\ 
      & & 3 & -0.461 & -0.467 & -0.470 & 0.031 & -0.531 & -0.412 \\ 
      & & 4 & -0.608 & -0.617 & -0.619 & 0.034 & -0.685 & -0.554 \\ 
      & & 5 & -0.721 & -0.699 & -0.696 & 0.065 & -0.808 & -0.561 \\ \cline{2-3}
      & \multirow{2}{*}{gender} & F & 0.535 & 0.540 & 0.541 & 0.066 & 0.407 & 0.679 \\ 
      & & M & 0.670 & 0.705 & 0.720 & 0.100 & 0.535 & 0.919 \\ \cline{2-3}
      & \multirow{8}{*}{deductible} & B & 1.337 & 1.383 & 1.408 & 0.216 & 1.036 & 1.866 \\ 
      & & C & 0.209 & 0.256 & 0.303 & 0.152 & 0.073 & 0.624 \\ 
      & & D & 0.603 & 0.627 & 0.672 & 0.161 & 0.397 & 0.982 \\ 
      & & E & 0.800 & 0.825 & 0.864 & 0.157 & 0.625 & 1.189 \\ 
      & & F & 0.753 & 0.774 & 0.819 & 0.155 & 0.592 & 1.154 \\ 
      & & G & 0.756 & 0.786 & 0.825 & 0.158 & 0.580 & 1.169 \\ 
      & & H & 0.820 & 0.829 & 0.824 & 0.190 & 0.463 & 1.149 \\ \cline{2-3}
      & \multirow{1}{*}{genderMarital} & B & -0.368 & -0.236 & -0.169 & 0.233 & -0.457 & 0.311 \\  \hline
      \multirow{17}{*}{$\bgamma$} & (Intercept) & -- & 6.423 & 6.429 & 6.415 & 0.073 & 6.310 & 6.504 \\ \cline{2-3}
      & \multirow{8}{*}{age.car} & 0 & -0.790 & -0.809 & -0.811 & 0.040 & -0.891 & -0.731 \\ 
      & & 1 & -1.006 & -1.025 & -1.028 & 0.039 & -1.099 & -0.954 \\ 
      & & 2 & -0.875 & -0.885 & -0.889 & 0.039 & -0.972 & -0.819 \\ 
      & & 3 & -0.842 & -0.861 & -0.863 & 0.039 & -0.950 & -0.799 \\ 
      & & 4 & -0.831 & -0.844 & -0.848 & 0.039 & -0.923 & -0.777 \\ 
      & & 5 & -0.781 & -0.792 & -0.795 & 0.039 & -0.879 & -0.728 \\ 
      & & 6 & -0.762 & -0.766 & -0.769 & 0.039 & -0.849 & -0.703 \\ 
      & & 7 & -0.725 & -0.732 & -0.735 & 0.039 & -0.814 & -0.662 \\ \cline{2-3}
      & \multirow{1}{*}{risk} & S & 0.110 & 0.110 & 0.110 & 0.012 & 0.087 & 0.132 \\ \cline{2-3} 
      & \multirow{1}{*}{agec} & 5 & -0.262 & -0.262 & -0.261 & 0.030 & -0.316 & -0.201 \\ \cline{2-3}
      &  \multirow{6}{*}{deductible} & B & -1.023 & -1.026 & -1.029 & 0.158 & -1.328 & -0.721 \\ 
      & & C & -0.534 & -0.553 & -0.571 & 0.069 & -0.700 & -0.461 \\ 
      & & D & -0.591 & -0.611 & -0.634 & 0.067 & -0.760 & -0.534 \\ 
      & & E & -0.533 & -0.552 & -0.575 & 0.066 & -0.702 & -0.478 \\ 
      & & F & -0.377 & -0.390 & -0.416 & 0.065 & -0.543 & -0.335 \\ 
      & & H & 0.352 & 0.361 & 0.369 & 0.107 & 0.170 & 0.591 \\  \hline
      $\xi$ & -- & & 1.667 & 1.667 & 1.667 & 0.001 & 1.665 & 1.670 \\
       \hline
       \hline
    \end{tabular}
    }
    \end{table}
    Comparing the models we observe that M4 produces the most optimal model fit criteria among the models considered. This extends to out-sample performance when predicting policy premiums. Plots produced for spatial effects in models M3 and M4 are mean adjusted. Since specification of M3 and M4 differ only in presence/absence of the hierarchical Bayesian variable selection component, the produced spatial effects mimic each other after adjusting for the mean. Comparing results in Tables \ref{tab::mc-ssdglm} and \ref{tab::mc-spssdglm} we observe that including the spatial effect results in more categories for vehicle age, driver age and deductible being selected. Overall, we observe that the findings remain consistent with our earlier research \citep[see,][]{halder2022spatial} but within a more robust proposed model choice framework. This is evident when comparing model estimates between Table \ref{tab::mc-spssdglm} and Table \ref{tab:mc-mean-spdglm-m3} and \ref{tab:mc-disp-spdglm-m3} 
    in the Supplement. The marital status and interaction between gender and marital status not being selected. We conclude by observing that the spatial effects are significantly positive in major cities in Connecticut indicating higher spatial risk as opposed to sparsely populated regions showing significantly lower risk.
    \begin{table}[t]
     \centering
	\caption{\small Table showing AIC and out-sample square root deviance for models M1--M4.}\label{tab:tab-res}
 \resizebox{\linewidth}{!}{
  \begin{tabular}{c|@{\extracolsep{60pt}}cccc@{}}
    \hline
    \hline
    \multirow{2}{*}{} & \multirow{2}{*}{M1} & \multirow{2}{*}{M2} & \multirow{2}{*}{M3} & \multirow{2}{*}{M4} \\ 
    &&&&\\
    \hline  
    \multirow{2}{*}{AIC} & \multirow{2}{*}{1340060} & \multirow{2}{*}{1117733} & \multirow{2}{*}{1117114} &\multirow{2}{*}{\bf 1115363}\\ 
    &&&&\\\hline
    \multirow{2}{*}{$\sqrt{\mbox{ Deviance }}$} & \multirow{2}{*}{5565.209} &\multirow{2}{*}{ 5509.549} & \multirow{2}{*}{5507.926} & \multirow{2}{*}{\bf 5441.23}\\
    &&&&\\
    \hline
    \hline
    \end{tabular}
 }
    \end{table}
    
    \section{Discussion}
     Double generalized linear models have not seen much use after their inception by \cite{lee2006double}. Hindrances presented by ambiguities existing around model specification/choice have been addressed in this paper. We propose Bayesian modeling frameworks that perform model selection using continuous spike and slab priors for hierarchical double generalized linear Tweedie spatial process models. Leveraging Langevin dynamics we are able to successfully produce practical implementations for the proposed frameworks which would otherwise remain unachievable with standard MCMC techniques. The proposed algorithms are available as a publicly accessible package for the \texttt{R}--statistical environment. Although the formulation considers the CP-g densities, evidently such modeling could be effected under any probabilistic framework that allows for varying dispersion. The application offers some key insights into the actuarial domain. It is generally believed that marital status and gender play a key role. However, the model inference suggests otherwise, with marital status not being selected as a significant feature.
     
     Future work is aimed at extending this framework in multiple directions. Firstly, with the advent of modern Bayesian variable selection priors---for example, the Bayesian Lasso, the Horseshoe prior etc., a comparative model selection performance remains to be seen when considered within hierarchical DGLM formulations. Secondly, with the emerging techniques for handling large spatial and spatiotemporal data \citep[see for e.g.,][]{heaton2019case} the DGLM framework could be extended to model spatially or spatio-temporally indexed observations over massive geographical domains. With respect to our application, this would allow us to investigate properties of the premium surface over much larger domains, for instance a country-wide study. Finally, extending these models to a spatiotemporal setting could be achieved using spatiotemporal covariance kernels that are commonly used. Depending on the nature of spatial and temporal interaction, we can have separable and non-separable kernels at our disposal \citep[see,][and references therein]{cressie2015statistics}. Bayesian variable selection could then be effected to examine resulting changes in model specification upon inclusion of random effects that address spatiotemporal variation in the data.

    \spacingset{1.0}
    \title{\bf  Supplementary Materials for\\ ``Bayesian Variable Selection in Double Generalized Linear Tweedie Spatial Process Models"}
		\maketitle
  \spacingset{1.45}
  \newpage
    \section{Posteriors}\label{sec:post}
    Under chosen priors on $\{\btheta_{m},\btheta_{pr}\}=\{\bbeta, \bgamma, \xi, \sigma^2,\phi_s\}$, the resulting joint posterior is specified by
	\begin{align}\label{eq:jposterior}
    	\begin{aligned}
    	 \pi\Big(\btheta_{m},\btheta_{pr}\mid \y\Big) &\propto U\Big(\xi\mid a_{\xi},b_{\xi}\Big)\times N_p\Big(\bbeta\mid \0_p,\blambda_\beta\trans\I_p\Big) \times N_q\Big(\bgamma\mid \0_q,\blambda_\gamma\trans\I_q\Big)\times\\
    	 &\hspace{.5cm}U\Big(\phi_s\mid a_{\phi_s}, b_{\phi_s}\Big)\times {\rm Gamma}\Big(\sigma^{-2}\mid a_{\sigma}, b_{\sigma}\Big)\times N_L\Big(\w\mid \0_L,\sigma^2\R(\phi_s)\Big)\times\\
    	 &\hspace{.5cm}{Tw}\Big(\y\mid \bmu=\exp(\X\bbeta+\F \w),\bphi=\exp(\Z\bgamma),\xi\Big),
    	\end{aligned}
	\end{align}
	We list the posteriors for individual parameters in the following equations, 
	\begin{equation}
	    \begin{aligned}\label{eq:posteriors-1a}
	        &\pi\left(\bbeta\mid \bgamma,\xi\right) \propto \exp\left\{-\left(\sum\limits_{i=1}^{L}\sum\limits_{j=1}^{n_i} \frac{1}{\phi_{ij}}d(y_{ij}\mid \mu_{ij}(\bbeta,\w),\xi)+\frac{\sigma^{-2}_{\beta}}{2}||\bbeta||_2^2+\frac{\sigma^{-2}}{2}\w\trans\R^{-1}(\phi_s)\w\right)\right\},\\
	        &\pi\left(\bgamma\mid \bbeta,\xi\right) \propto  \exp\left\{-\left(\sum\limits_{i=1}^{L}\sum\limits_{j=1}^{n_i} \frac{1}{\phi_{ij}( \bgamma)} d(y_{ij}\mid \mu_{ij},\xi)+\frac{\log\phi_{ij}(\bgamma)}{2}I(y_{ij}>0)+\frac{\sigma^{-2}_{\gamma}}{2}||\bgamma||_2^2\right)\right\},\\
	        &\pi(\xi\mid \bbeta,\bgamma)\propto \prod_{i=1}^{L}\prod_{j=1}^{n_i}c_{ij}(y_{ij}\mid \phi_{ij},\xi)\exp\left\{- \frac{1}{\phi_{ij}}d(y_{ij}\mid \mu_{ij},\xi)\right\}I\left(\xi\in (a_{\xi},b_{\xi})\right),\\
	        &\pi(\phi_s) \propto |\R(\phi_s)|^{-1/2}\exp\left(-\frac{\sigma^{-2}}{2}\w\trans\R^{-1}(\phi_s)\w\right)I\left(\phi_s\in (a_{\phi_s},b_{\phi_s})\right),\\ 
	        &\pi(\sigma^{-2}) = {\rm Gamma}\left(a_{\sigma}+\frac{L}{2}, b_{\sigma}+\frac{1}{2}\w\trans\R^{-1}(\phi_s)\w\right),
	    \end{aligned}
	\end{equation}
	where $|\cdot|$ denotes the determinant and $d(y_{ij}\mid \mu_{ij},\xi)$ is the deviance function. In the scenario where we have no spatial effect in the DGLM, the reduced set of parameters are $\btheta_{m}=\{\bbeta,\bgamma,\xi\}$ having a similar joint posterior 
	after omitting the second row involving spatial process prior specification and setting $\bmu=\exp(\X\bbeta)$ in the likelihood, keeping the same prior specifications on the other parameters, the resulting posteriors are as follows,
    \begin{align*}
	        &\pi\left(\bbeta\mid \bgamma,\xi\right) \propto \exp\left\{-\left(\sum\limits_{i=1}^{N} \frac{1}{\phi_{i}}d(y_{i}\mid \mu_{i}(\bbeta),\xi)+\frac{\sigma^{-2}_{\beta}}{2}||\bbeta||_2^2\right)\right\},\\
	        &\pi\left(\bgamma\mid \bbeta,\xi\right) \propto  \exp\left\{-\left(\sum\limits_{i=1}^{N} \frac{1}{\phi_{i}( \bgamma)} d(y_{i}\mid \mu_{i},\xi)+\frac{1}{2}\log\phi_{ij}(\bgamma)I(y_{ij}>0)+\frac{\sigma^{-2}_{\gamma}}{2}||\bgamma||_2^2\right)\right\},\\
	        & \pi(\xi)\propto \prod_{i=1}^{N}c_{i}(y_{i}\mid \phi_{i},\xi)\exp\left\{- \frac{1}{\phi_{i}}d(y_{i}\mid \mu_{i},\xi)\right\}I\left(\xi\in (a_{\xi},b_{\xi})\right).
	\end{align*}
	Updates leveraging MALA for ${\bbeta, \w}$ (or $\bbeta$) and $\bgamma$ require the proposals to be specified appropriately using the gradients of log-posterior densities, $\nabla\log\pi\left(\bbeta\mid \bgamma,\xi\right)$ and $\nabla\log\pi\left(\bgamma\mid \bbeta,\xi\right)$ respectively. Candidate samples are obtained using,
	\begin{equation}
	    \begin{aligned}\label{eq:candidates}
	        &{(\bbeta, \w)\trans}^{*}=(\bbeta,\w)\trans+\frac{\tau_{\beta,w}^2}{2}\A_{\beta,w}\nabla\log\pi\left(\bbeta\mid \bgamma,\xi\right)+\tau_{\beta,w}\A_{\beta,w}^{1/2}\cdot N_{p+L}(\0,\I_{p+L}),\\
	        &\bgamma^{*}=\bgamma+\frac{\tau_{\gamma}^2}{2}\A_{\gamma}\nabla\log\pi\left(\gamma\mid \bbeta,\xi\right)+\tau_{\gamma}\A_{\gamma}^{1/2}\cdot N_{q}(\0,\I_{q}),
	    \end{aligned}
	\end{equation}
	where $\A_{\beta,w}^{-1}={\rm E}\left[-\nabla^2\log\pi\left(\bbeta_{\w}\mid -\right)\right]$ and $\A_{\gamma}^{-1}={\rm E}\left[-\nabla^2\log\pi\left(\bgamma\mid -\right)\right]$.

	Under a continuous spike and slab prior on $\bbeta$ and $\bgamma$ the posteriors for the hierarchical latent DGLM is as follows,
	\begin{equation}
    	\begin{aligned}\label{eq:posterior-1b}
    	    &\pi\left(\bbeta_{\w}\mid -\right) \propto \exp\left\{-\left(\sum\limits_{i=1}^{L}\sum\limits_{j=1}^{n_i} \frac{1}{\phi_{ij}}d(y_{ij}\mid \mu_{ij}(\bbeta,\w),\xi)+\frac{1}{2}\bbeta\trans\bGamma_{\beta}^{-1}\bbeta+\frac{\sigma^{-2}}{2}\w\trans\R^{-1}(\phi_s)\w\right)\right\}\\
    	    &\left. \begin{array}{l}
    	    \bGamma_{\beta}={\rm diag}\{\zeta_{\beta,i_b}\sigma_{\beta,i_b}^2\}\\
    	    \pi(\zeta_{\beta,i_b}\mid -) = \frac{\alpha_{1,\beta}}{\alpha_{1,\beta}+\alpha_{2,\beta}}\delta_{\nu_0}(\cdot)+\frac{\alpha_{2,\beta}}{\alpha_{1,\beta}+\alpha_{2,\beta}}\delta_{\nu_1}(\cdot)\\
    	    \pi(\sigma^{-2}_{\beta,i_b}\mid -) = {\rm Gamma}\left(a_\sigma+\frac{1}{2},b_{\sigma}+\frac{\beta^2_{i_b}}{2\zeta_{\beta,i_b}}\right)\\
    	    \pi(\alpha_{\beta}\mid -) = {\rm Beta}\left(1+\#\{i_b:\zeta_{\beta,i_b}=1\},1+\#\{i_b:\zeta_{\beta,i_b}=\nu_0\}\right)\end{array}\right\}, i_b = 1,2,\ldots,p
    	    \\
    	    &\pi\left(\bgamma\mid -\right) \propto  \exp\left\{-\left(\sum\limits_{i=1}^{L}\sum\limits_{j=1}^{n_i} \frac{1}{\phi_{ij}( \bgamma)} d(y_{ij}\mid \mu_{ij},\xi)+\frac{1}{2}\log\phi_{ij}(\bgamma)I(y_{ij}>0)+\frac{1}{2}\bgamma\trans\bGamma_{\gamma}^{-1}\bgamma\right)\right\},\\
	        &\left. \begin{array}{l}
	        \bGamma_{\gamma}={\rm diag}\{\zeta_{\gamma,i_g}\sigma_{\gamma,i_g}^2\}\\
    	    \pi(\zeta_{\gamma,i_g}\mid -) = \frac{\alpha_{1,\gamma}}{\alpha_{1,\gamma}+\alpha_{2,\gamma}}\delta_{\nu_0}(\cdot)+\frac{\alpha_{2,\gamma}}{\alpha_{1,\gamma}+\alpha_{2,\gamma}}\delta_{\nu_1}(\cdot)\\
    	    \pi(\sigma^{-2}_{\gamma,i_g}\mid -) ={\rm Gamma}\left(a_\sigma+\frac{1}{2},b_{\sigma}+\frac{\gamma^2_{i_g}}{2\zeta_{\gamma,i_g}}\right)\\
    	    \pi(\alpha_{\gamma}\mid -) = {\rm Beta}\left(1+\#\{i_g:\zeta_{\gamma,i_g}=1\},1+\#\{i_g:\zeta_{\gamma,i_g}=\nu_0\}\right)\end{array}\right\},i_g = 1,2,\ldots,q
    	    \\
	        &\pi(\xi\mid -)\propto \prod_{i=1}^{L}\prod_{j=1}^{n_i}c_{ij}(y_{ij}\mid \phi_{ij},\xi)\exp\left\{- \frac{1}{\phi_{ij}}d(y_{ij}\mid \mu_{ij},\xi)\right\}I\left(\xi\in (a_{\xi},b_{\xi})\right),\\
	        &\pi(\phi_s\mid -) \propto |\R(\phi_s)|^{-1/2}\exp\left(-\frac{\sigma^{-2}}{2}\w\trans\R^{-1}(\phi_s)\w\right)I\left(\phi_s\in (a_{\phi_s},b_{\phi_s})\right),\\ 
	        &\pi(\sigma^{-2}\mid -) = {\rm Gamma}\left(a_{\sigma}+\frac{L}{2}, b_{\sigma}+\frac{1}{2}\w\trans\R^{-1}(\phi_s)\w\right),
    	\end{aligned}
	\end{equation}
	where,
	\begin{equation*}
	    \begin{aligned}
	        &\alpha_{1,\beta}=(1-\alpha_\beta)\nu_0^{-1/2}\exp\left\{-\frac{\beta_{i_b}^2}{2\nu_0\sigma_{\beta,i_b}^2}\right\}, ~ \alpha_{2,\beta}=\alpha_\beta\exp\left\{-\frac{\beta_{i_b}^2}{2\sigma^2_{\beta,i_b}}\right\}, i_b = 1,2,\ldots,p;\\
	        &\alpha_{1,\gamma}=(1-\alpha_\gamma)\nu_0^{-1/2}\exp\left\{-\frac{\gamma_{i_g}^2}{2\nu_0\sigma_{\gamma,i_g}^2}\right\}, ~ \alpha_{2,\gamma}=\alpha_\gamma\exp\left\{-\frac{\gamma_{i_g}^2}{2\sigma^2_{\gamma,i_g}}\right\}, i_g = 1,2,\ldots,q.
	    \end{aligned}
	\end{equation*}
	Under no latent specification the resulting posteriors can be obtained similarly, by omitting updates for the latent process and associated process parameters/hyper-parameters. The joint posterior is sampled by leveraging the posteriors in eq. (\ref{eq:posterior-1b}) and employing a similar hybrid sampling strategy as earlier, with additional Gibbs updates for the spike and slab parameters, $\{\bzeta_\beta,\bsigma^2_\beta, \alpha_\beta, \bzeta_\gamma,\bsigma^2_\gamma, \alpha_\gamma\}$.
	
	\section{Experiments, Diagnostics and Software}\label{sec:ede}
        
        We outline supporting simulation experiments, convergence diagnostics and details on available software to supplement Section 3 in the manuscript. We begin with results from additional experiments.
        \subsection{Supporting Experiments}
        We performed simulation experiments to assess performance of model M2---Table \ref{tab:synexp-1} shows the results featuring the same settings as outlined in Section 3, only without the spatial effects.
        \begin{table}[H]
        \centering
        \resizebox{\linewidth}{!}{
            \begin{tabular}{c|c|@{\extracolsep{10em}}cccc@{}}
          \hline
          \hline
          \multirow{2}{*}{Configuration} & \multirow{2}{*}{Prop. of 0s} & \multirow{2}{*}{$CP(\btheta)$} & \multirow{2}{*}{False Positive Rate} & \multirow{2}{*}{True Positive Rate}\\
          &&&&\\
          \hline
          \multirow{4}{*}{Configuration 1} & 0.15 & 1.00 (0.00) & 0.00 (0.00) & 1.00 (0.00) \\ 
          & 0.30 & 1.00 (0.00) & 0.00 (0.00) & 1.00 (0.00) \\ 
          & 0.60 & 1.00 (0.00) & 0.00 (0.00) & 1.00 (0.00) \\ 
          & 0.80 & 1.00 (0.10) & 0.04 (0.06) & 0.99 (0.03) \\\hline
          \multirow{4}{*}{Configuration 2} & 0.15 & 1.00 (0.00) & 0.01 (0.05) & 0.99 (0.03) \\ 
          & 0.30 & 1.00 (0.00) & 0.00 (0.00) & 1.00 (0.00) \\ 
          & 0.60 & 1.00 (0.10) & 0.00 (0.00) & 0.98 (0.04) \\ 
          & 0.80 & 0.90 (0.10) & 0.01 (0.05) & 0.99 (0.03) \\ \hline
          \multirow{4}{*}{Configuration 3} & 0.15 & 0.90 (0.10) & 0.00 (0.00) & 1.00 (0.00) \\ 
          & 0.30 & 1.00 (0.00) & 0.01 (0.04) & 1.00 (0.00) \\ 
          & 0.60 & 1.00 (0.00) & 0.00 (0.00) & 0.99 (0.03) \\ 
          & 0.80 & 1.00 (0.00) & 0.05 (0.09) & 0.96 (0.05) \\ 
          \hline
          \hline
        \end{tabular}
        }
        \caption{Results for synthetic experiments corresponding to Table 3 for model M2 showing average coverage probabilities for the estimated model coefficients.}\label{tab:synexp-11}
    \end{table}
    In what follows, we outline simulation experiments for the scenario where we have spatial covariates in our data. Finally, we focus on showing simulations results for different synthetic spatial patterns.
    \subsubsection{Spatial Covariates}
    Setting up experiments with spatial covariates we are mindful of possible endogeneity issues arising from included spatial covariates being correlated with the true spatial pattern \citep[see, for e.g.,][]{fan2014endogeneity}. We use the following settings and true values: $N=1\times 10^4$, $L=1\times 10^2$, $\bbeta = (1.0, 1.5, 1\times 10^{-5}, 1.4, 1.1, 1\times 10^{-5}, 2.5)\trans$, $\bgamma = (1.0, 1\times 10^{-5},  1.5, 1.1, 1\times 10^{-5}, -2.5, 1\times 10^{-5})\trans$. In $\X$ and $\Z$ we include two spatial covariates in the last two columns: $\x_6=\z_6\sim N(5(\cos(3\pi s_x)+\sin(3\pi s_y)), 1)$ and $\x_7=\z_7\sim N(2(\cos(\pi s_x)+\sin(\pi s_y)), 1)$, while the rest were sampled from standard Gaussian distributions resulting in an average of 50\% zeros in the data. Under this setup, we observe that for the mean model $\x_6$ is significant and $\x_7$ is not while for the dispersion model it is the other way around. The true spatial effect is simulated using $\sigma^2=1.5$ and $\phi_s=3$ from an exponential kernel. The true index parameter, $\xi=1.5$. The design matrices were centered and scaled. The resulting absolute value of the correlations within covariates and between them and the true spatial effect from the setup above was $<0.01$. To assess the performance of our models we focus on M3 and and M4 and compute the MSE, coverage probability, FPR and TPR for each of these models. We performed 100 replications under these settings. The results are shown below in Table \ref{tab:sim-spcov}. Models M3 and M4 were able to estimate coefficients $\beta_6$, $\beta_7$ and $\gamma_6$, $\gamma_7$ with sufficient accuracy.
   \begin{table}[ht]
    \centering
         \begin{tabular}{c|cccccc}
      \hline
      \hline
       Model & MSE ($\btheta$) & MSE ($\w$) & CP ($\btheta$) & CP ($\w$) & FPR & TPR \\ 
      \hline
      \multirow{2}{*}{M3} & 0.008 & 0.072 & 0.907 & 0.940 & -- & -- \\ 
      & (0.019) & (0.186) & (0.111) & (0.092) & -- & -- \\ 
       \hline
      \multirow{2}{*}{M4} & 0.003 & 0.062 & 0.916 & 0.959 & 0.000 & 1.000 \\ 
      & (0.004) & (0.129) & (0.063) & (0.09) & (0.165) & (0.035) \\ 
      \hline
      \hline
    \end{tabular}
    \caption{Table showing results for simulation carried using spatial covariates in design matrices $\X$ and $\Z$.}\label{tab:sim-spcov}
    \end{table}
    \subsubsection{Spatial Patterns}
    We show the results for simulation experiments performed for patters shown in Figure 1 of the manuscript in Tables \ref{tab:sim-pat-1} and \ref{tab:sim-pat-2} below.
	
    \begin{table}[H]
    \centering
    \resizebox{\linewidth}{!}{
        \begin{tabular}{c|c|@{\extracolsep{6em}}cccccccc@{}}
    \hline
    \hline
    \multirow{2}{*}{$N$} & \multirow{2}{*}{Prop. of 0s} & \multicolumn{6}{c}{MSE} & \multicolumn{2}{c}{CP($\btheta$)}\\ 
    \cline{3-8}\cline{9-10}
    && $\bbeta$ & $\bgamma$ & $\w$ & $\xi$ & $\sigma^2$ & $\phi_s$ & $\btheta_{m}$ & $\btheta_{pr}$\\
      \hline
      \multirow{8}{*}{2000}& 0.15 & 0.00 & 0.00 & 0.02 & 0.00 & 0.04 & 1.45 & 1.00 & 0.95 \\ 
      & (0.06) & (0.00) & (0.00) & (0.01) & (0.00) & (0.05) & (2.98) & (0.00) & (0.01) \\
      \cline{2-2}
      & 0.30 & 0.00 & 0.00 & 0.06 & 0.00 & 0.11 & 4.48 & 1.00 & 0.94 \\ 
      & (0.07) & (0.00) & (0.00) & (0.02) & (0.00) & (0.07) & (6.12) & (0.00) & (0.03) \\ 
      \cline{2-2}
      & 0.60 & 0.00 & 0.00 & 0.08 & 0.00 & 0.08 & 6.46 & 0.98 & 0.94 \\ 
      & (0.05) & (0.00) & (0.00) & (0.02) & (0.00) & (0.12) & (14.51) & (0.05) & (0.03) \\ 
      \cline{2-2}
      & 0.80 & 0.01 & 0.00 & 0.21 & 0.00 & 0.09 & 10.00 & 0.96 & 0.92 \\ 
      & (0.04) & (0.00) & (0.00) & (0.04) & (0.00) & (0.08) & (21.00) & (0.06) & (0.03) \\\hline 
      \multirow{8}{*}{5000}& 0.15 & 0.00 & 0.00 & 0.01 & 0.00 & 0.06 & 1.89 & 0.99 & 0.93 \\ 
      & (0.05) & (0.00) & (0.00) & (0.00) & (0.00) & (0.08) & (3.07) & (0.04) & (0.02) \\ 
      \cline{2-2}
      & 0.30 & 0.00 & 0.00 & 0.03 & 0.00 & 0.09 & 5.07 & 0.94 & 0.93 \\ 
      & (0.05) & (0.00) & (0.00) & (0.01) & (0.00) & (0.08) & (9.23) & (0.11) & (0.02) \\ 
      \cline{2-2}
      & 0.60 & 0.00 & 0.00 & 0.04 & 0.00 & 0.04 & 2.05 & 0.99 & 0.94 \\ 
      & (0.07) & (0.00) & (0.00) & (0.01) & (0.00) & (0.08) & (4.86) & (0.04) & (0.02) \\ 
      \cline{2-2}
      & 0.80 & 0.00 & 0.00 & 0.10 & 0.00 & 0.10 & 12.86 & 0.98 & 0.94 \\ 
      & (0.05) & (0.00) & (0.00) & (0.03) & (0.00) & (0.08) & (27.54) & (0.05) & (0.02) \\\hline 
      \multirow{8}{*}{10000}& 0.15 & 0.00 & 0.00 & 0.00 & 0.00 & 0.13 & 1.30 & 0.99 & 0.95 \\
      & (0.05) & (0.00) & (0.00) & (0.00) & (0.00) & (0.21) & (2.72) & (0.04) & (0.01) \\ 
      \cline{2-2}
      & 0.30 & 0.00 & 0.00 & 0.01 & 0.00 & 0.10 & 2.70 & 0.99 & 0.94 \\ 
      & (0.08) & (0.00) & (0.00) & (0.00) & (0.00) & (0.08) & (3.88) & (0.04) & (0.03) \\ 
      \cline{2-2}
      & 0.60 & 0.00 & 0.00 & 0.02 & 0.00 & 0.07 & 4.16 & 0.95 & 0.95 \\ 
      & (0.06) & (0.00) & (0.00) & (0.00) & (0.00) & (0.09) & (9.62) & (0.06) & (0.02) \\ 
      \cline{2-2}
      & 0.80 & 0.00 & 0.00 & 0.06 & 0.00 & 0.05 & 1.34 & 0.99 & 0.93 \\ 
      & (0.04) & (0.00) & (0.00) & (0.02) & (0.00) & (0.04) & (1.61) & (0.04) & (0.03) \\ 
    \hline
    \hline
    \end{tabular}
    }
    \caption{Results for Pattern 1, with number of locations, $L = 100$.}\label{tab:sim-pat-1}
    \end{table}
    
    \begin{table}[H]
    \centering
    \resizebox{\linewidth}{!}{
        \begin{tabular}{c|c|@{\extracolsep{10em}}cccccc@{}}
    \hline
    \hline
    \multirow{2}{*}{$N$} & \multirow{2}{*}{Prop. of 0s} & \multicolumn{4}{c}{MSE} & \multicolumn{2}{c}{CP($\btheta$)}\\ 
    \cline{3-6}\cline{7-8}
    && $\bbeta$ & $\bgamma$ & $\w$ & $\xi$  & $\btheta_{m}$ & $\btheta_{pr}$\\
      \hline
      \multirow{8}{*}{2000} & 0.15 & 0.00 & 0.00 & 0.06 & 0.00  & 0.98 & 0.93 \\ 
      & (0.01) & (0.00) & (0.00) & (0.03) & (0.00) & (0.05) & (0.03) \\ 
      \cline{2-2}
      & 0.30 & 0.00 & 0.00 & 0.34 & 0.00 &  0.96 & 0.95 \\ 
      & (0.01) & (0.00) & (0.00) & (0.10) & (0.00) &  (0.06) & (0.02) \\
      \cline{2-2}
      & 0.60 & 0.00 & 0.00 & 0.92 & 0.00 &  0.96 & 0.94 \\ 
      & (0.01) & (0.00) & (0.00) & (0.31) & (0.00) &  (0.08) & (0.03) \\ 
      \cline{2-2}
      & 0.80 & 0.00 & 0.00 & 2.74 & 0.00 &  1.00 & 0.91 \\ 
      & (0.01) & (0.00) & (0.00) & (0.26) & (0.00) &  (0.00) & (0.02) \\\hline 
      \multirow{8}{*}{5000} & 0.15 & 0.00 & 0.00 & 0.03 & 0.00 &  0.95 & 0.95 \\ 
      & (0.01) & (0.00) & (0.00) & (0.02) & (0.00) &  (0.11) & (0.02) \\ 
      \cline{2-2}
      & 0.30 & 0.00 & 0.00 & 0.13 & 0.00 &  0.98 & 0.93 \\ 
      & (0.01) & (0.00) & (0.00) & (0.04) & (0.00) &  (0.05) & (0.03) \\ 
      \cline{2-2}
      & 0.60 & 0.00 & 0.00 & 0.46 & 0.00 &  1.00 & 0.95 \\ 
      & (0.01) & (0.00) & (0.00) & (0.15) & (0.00) &  (0.00) & (0.02) \\ 
      \cline{2-2}
      & 0.80 & 0.00 & 0.00 & 1.68 & 0.00 &  0.91 & 0.91 \\ 
      & (0.01) & (0.00) & (0.00) & (0.28) & (0.00) &  (0.10) & (0.03) \\\hline 
      \multirow{8}{*}{10000}& 0.155 & 0.00 & 0.00 & 0.01 & 0.00 & 0.92 & 0.93 \\ 
      & (0.01) & (0.00) & (0.00) & (0.01) & (0.00) & (0.06) & (0.03) \\ 
      \cline{2-2}
      & 0.30 & 0.00 & 0.00 & 0.09 & 0.00 &  0.99 & 0.93 \\ 
      & (0.01) & (0.00) & (0.00) & (0.03) & (0.00) & (0.04) & (0.03) \\
      \cline{2-2}
      & 0.60 & 0.00 & 0.00 & 0.30 & 0.00 & 0.96 & 0.93 \\ 
      & (0.01) & (0.00) & (0.00) & (0.07) & (0.00) & (0.06) & (0.02) \\ 
      \cline{2-2}
      & 0.80 & 0.00 & 0.00 & 1.37 & 0.00 & 0.98 & 0.93 \\ 
      & (0.01) & (0.00) & (0.00) & (0.19) & (0.00) & (0.05) & (0.03) \\ 
    \hline
    \hline
    \end{tabular}
    }
    \caption{Results for Pattern 2, with number of locations, $L = 100$.}\label{tab:sim-pat-2}
    \end{table}
    \subsection{Convergence}\label{subsec:conv}
    To assess convergence we resort to monitoring trace plots, posterior density and auto-correlation (ACF) plots for model parameters and coefficients. We show them particularly for the index parameter $\xi$ in Figure \ref{fig:xi-mcmc} below. They were produced using parameter settings outlined in the examples within the \texttt{R}-package. Additional plots for $\bbeta$, $\bgamma$, $\sigma^2$ and $\phi_s$ are included in examples within the \texttt{R}-package described in the next subsection.
    \begin{figure}[h]
    \centering
    \includegraphics[width=\textwidth]{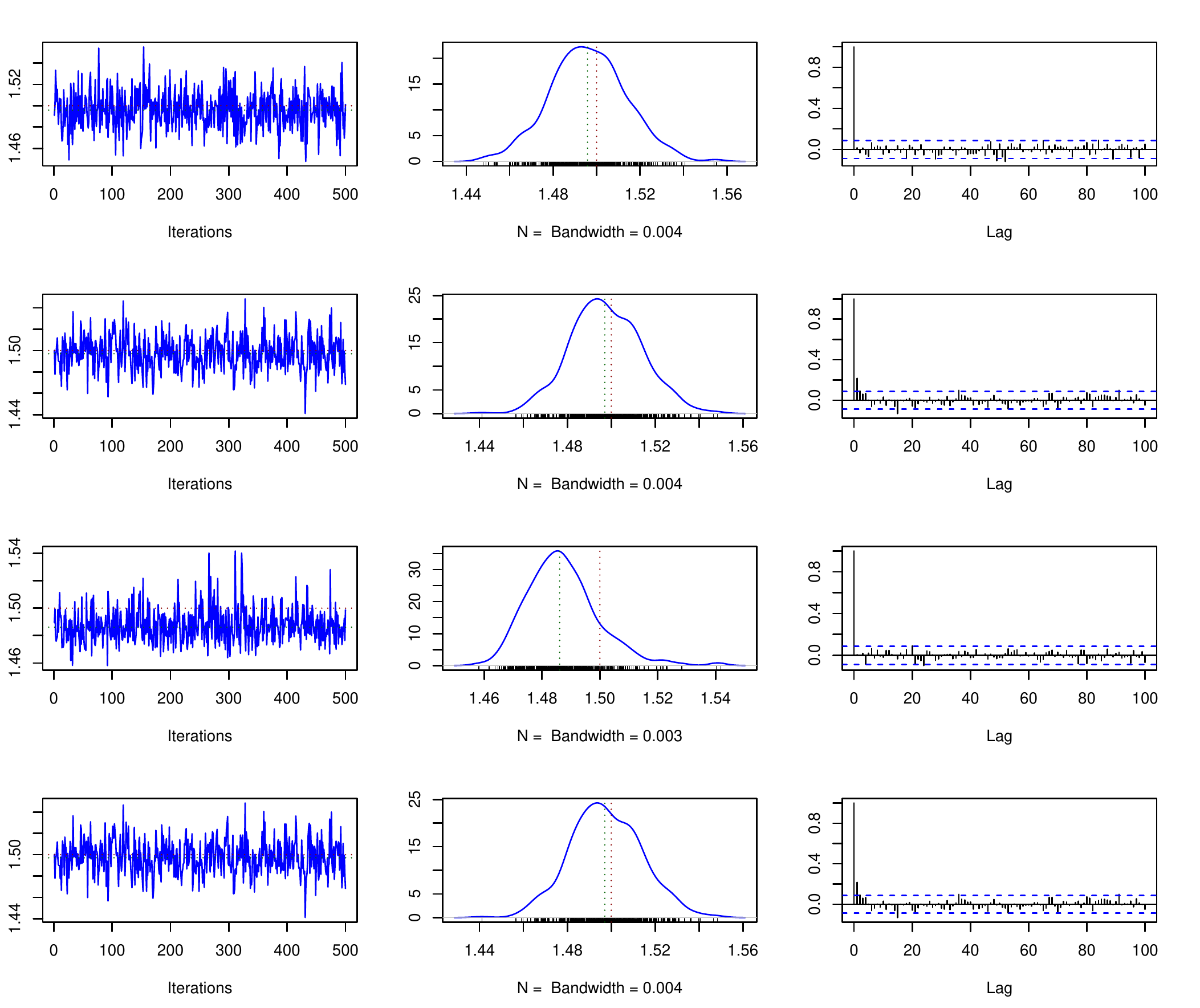}
    \caption{\small Convergence diagnostics for posterior samples of the index parameter, $\xi$. Each row corresponds to a model---(first row) M1 (second row) M2 (third row) M3 (fourth row) M4. In each row we show (left) trace, (center) posterior probability density and (right) ACF plots.}\label{fig:xi-mcmc}
    \end{figure}
    
    \subsection{R-package: \texttt{sptwdglm}}\label{subsec:r-pack}
    Algorithms featuring posteriors in Section \ref{sec:post} constitute our implementation that is available in the publicly accessible open-source repository: \url{https://github.com/arh926/sptwdglm}. It is composed of four Markov chain Monte Carlo samplers as described below. Each function is accompanied with examples in the package.
    \begin{enumerate}
        \item \texttt{dglm.autograd.R:} Fits a DGLM to the data. It requires the response as a vector, $y$, covariates as matrices $\X$ and $\Z$, upper and lower bounds for the index parameter $\xi$ as input at the minimum to produce an output. Other optional parameters are also provided for custom fine-tuning which have optimized defaults for ease of use. This is model M1 in the manuscript.
        \item \texttt{ssdglm.autograd.R:} Fits a DGLM featuring variable selection via the spike and slab prior to the data. It requires the response as a vector, $y$, covariates as matrices $\X$ and $\Z$, upper and lower bounds for the index parameter $\xi$ as input at the minimum to produce an output. Other optional parameters are also provided for custom fine-tuning which have optimized defaults for ease of use. This is model M2 in the manuscript.
        \item \texttt{spdglm.autograd.R:} Fits a spatial DGLM to the data. It requires the coordinates as a matrix, the response as a vector, $y$, covariates as matrices $\X$ and $\Z$, upper and lower bounds for the index parameter $\xi$ as input at the minimum to produce an output. Other optional parameters are also provided for custom fine-tuning which have optimized defaults for ease of use. This is model M3 in the manuscript.
        \item \texttt{spssdglm.autograd.R:} Fits a spatial DGLM featuring variable selection via the spike and slab prior to the data. It requires the coordinates as a matrix, the response as a vector, $y$, covariates as matrices $\X$ and $\Z$, upper and lower bounds for the index parameter $\xi$ as input at the minimum to produce an output. Other optional parameters are also provided for custom fine-tuning which have optimized defaults for ease of use. This is model M4 in the manuscript.
        \item \texttt{FDR.R:} Computes local false discovery rates (see page 10 of manuscript) for posterior samples of $\bbeta$ and $\bgamma$ arising from M2 or M4. Optional parameters to be specified are threshold defaulting at 0.05 and percentage at which FDR is computed defaulting to 5\%.
    \end{enumerate}
    
    \section{Automobile Insurance Premiums, Connecticut, 2008: Additional results}\label{sec:supp-real}
\begin{table}[H]
\centering
\resizebox{\linewidth}{!}{
    \begin{tabular}{c|c|@{\extracolsep{14em}}cccccc@{}}
  \hline
  \hline
  Parameters & Levels & Median & Mean & SD & Lower HPD & Upper HPD\\ 
  \hline
  (Intercept) & -- & 5.039 & 4.994 & 0.372 & 4.286 & 5.676\\ \cline{1-2}
  \multirow{8}{*}{age.car} & 0 & -0.373 & -0.344 & 0.224 & -0.787 & 0.111 \\ 
  & 1 & -0.306 & -0.271 & 0.218 & -0.641 & 0.224 \\ 
  & 2 & -0.196 & -0.164 & 0.219 & -0.519 & 0.333 \\ 
  & 3 & -0.371 & -0.331 & 0.262 & -0.740 & 0.262 \\ 
  & 4 & -0.187 & -0.156 & 0.225 & -0.528 & 0.369 \\ 
  & 5 & -0.111 & -0.078 & 0.227 & -0.454 & 0.437 \\ 
  & 6 & -0.350 & -0.320 & 0.230 & -0.696 & 0.213 \\ 
  & 7 & -0.401 & -0.367 & 0.252 & -0.855 & 0.138 \\ \cline{1-2}
  \multirow{1}{*}{risk} & S & -0.176 & -0.173 & 0.060 & -0.292 & -0.054 \\ \cline{1-2}
  \multirow{4}{*}{agec} & 2 & -0.014 & -0.013 & 0.097 & -0.208 & 0.163 \\ 
  & 3 & -0.292 & -0.292 & 0.101 & -0.504 & -0.105 \\ 
  & 4 & -0.150 & -0.152 & 0.114 & -0.347 & 0.096 \\ 
  & 5 & -0.430 & -0.424 & 0.230 & -0.881 & -0.022 \\ \cline{1-2}
  \multirow{2}{*}{gender} & F & 0.451 & 0.436 & 0.220 & 0.015 & 0.838 \\ 
  & M & 0.279 & 0.268 & 0.219 & -0.141 & 0.668 \\ \cline{1-2}
  \multirow{2}{*}{marital} & M & -4.096 & -3.941 & 1.031 & -5.532 & -1.829 \\ 
  & S & 4.954 & 4.893 & 1.751 & 2.285 & 7.588 \\  \cline{1-2}
  \multirow{7}{*}{deductible} & B & 2.613 & 2.595 & 0.832 & 0.987 & 4.065 \\ 
  & C & 1.025 & 0.999 & 0.439 & 0.049 & 1.779 \\ 
  & D & 1.519 & 1.487 & 0.432 & 0.605 & 2.322 \\ 
  & E & 1.626 & 1.579 & 0.429 & 0.651 & 2.375 \\ 
  & F & 1.691 & 1.678 & 0.424 & 0.695 & 2.412 \\ 
  & G & 1.733 & 1.717 & 0.433 & 0.879 & 2.626 \\ 
  & H & 1.804 & 1.754 & 0.630 & 0.552 & 2.885 \\ \cline{1-2}
  \multirow{4}{*}{genderMarital} & A & -4.998 & -4.940 & 1.725 & -7.626 & -2.515 \\ 
  & B & 3.756 & 3.622 & 1.057 & 1.413 & 5.219 \\ 
  & C & -4.521 & -4.532 & 1.823 & -7.360 & -1.766 \\ 
  & D & 3.850 & 3.685 & 0.986 & 1.814 & 5.412 \\  \hline
  \hline
\end{tabular}
}
\caption{\small Model coefficients for fixed effects in the mean model for M1. We show the median, mean, standard deviation and HPDs.}\label{tab:mc-mean-dglm-m1} 
\end{table}
\begin{table}[H]
\centering
\resizebox{\linewidth}{!}{
    \begin{tabular}{c|c|@{\extracolsep{14em}}cccccc@{}}
  \hline
  \hline
  Parameters & Levels & Median & Mean & SD & Lower HPD & Upper HPD\\ 
  \hline 
  (Intercept) & -- & 5.871 & 5.867 & 0.052 & 5.784 & 5.948 \\ \cline{1-2}
  \multirow{8}{*}{age.car} & 0 & -0.814 & -0.804 & 0.088 & -0.915 & -0.663 \\ 
  & 1 & -1.494 & -1.471 & 0.136 & -1.657 & -1.217 \\ 
  & 2 & -1.122 & -1.112 & 0.118 & -1.299 & -0.918 \\ 
  & 3 & 0.720 & 0.737 & 0.162 & 0.485 & 1.009 \\ 
  & 4 & -0.827 & -0.826 & 0.072 & -0.937 & -0.709 \\ 
  & 5 & -0.045 & -0.033 & 0.122 & -0.203 & 0.170 \\ 
  & 6 & -0.502 & -0.489 & 0.057 & -0.571 & -0.392 \\ 
  & 7 & 0.672 & 0.673 & 0.128 & 0.485 & 0.887 \\ \cline{1-2}
  \multirow{1}{*}{risk} & S & 1.688 & 1.693 & 0.048 & 1.620 & 1.772 \\ \cline{1-2}
  \multirow{4}{*}{agec} & 2 & 1.333 & 1.343 & 0.025 & 1.317 & 1.400 \\ 
  & 3 & 1.384 & 1.388 & 0.010 & 1.374 & 1.409 \\ 
  & 4 & 1.394 & 1.396 & 0.013 & 1.370 & 1.417 \\ 
  & 5 & 1.157 & 1.160 & 0.124 & 0.981 & 1.351 \\  \cline{1-2}
  \multirow{2}{*}{gender} & F & 0.168 & 0.151 & 0.053 & 0.055 & 0.214 \\ 
  & M & -0.074 & -0.075 & 0.027 & -0.118 & -0.024 \\   \cline{1-2}
  \multirow{2}{*}{marital} & M & 0.022 & 0.044 & 0.055 & -0.034 & 0.131 \\ 
  & S & 2.308 & 2.277 & 0.243 & 1.829 & 2.665 \\   \cline{1-2}
  \multirow{7}{*}{deductible} & B & -0.301 & -0.280 & 0.052 & -0.349 & -0.184 \\ 
  & C & -0.122 & -0.131 & 0.093 & -0.306 & -0.011 \\ 
  & D & -0.519 & -0.509 & 0.154 & -0.746 & -0.241 \\ 
  & E & -0.297 & -0.287 & 0.183 & -0.563 & 0.017 \\ 
  & F & 0.446 & 0.444 & 0.051 & 0.358 & 0.515 \\ 
  & G & 1.664 & 1.662 & 0.039 & 1.596 & 1.720 \\ 
  & H & 0.513 & 0.516 & 0.042 & 0.437 & 0.583 \\ \cline{1-2}
  \multirow{4}{*}{genderMarital} & A & -0.912 & -0.904 & 0.166 & -1.175 & -0.624 \\ 
  & B & 0.270 & 0.270 & 0.005 & 0.257 & 0.279 \\ 
  & C & -0.660 & -0.654 & 0.126 & -0.856 & -0.456 \\ 
  & D & -0.222 & -0.236 & 0.042 & -0.309 & -0.176 \\ \hline
  $\xi$ & -- & 1.674 & 1.674 & 0.007 & 1.664 & 1.687 \\
   \hline
   \hline
\end{tabular}
}
\caption{\small Model coefficients for fixed effects in the dispersion model for M1. We show the median, mean, standard deviation and HPDs.}\label{tab:mc-disp-dglm-m1}
\end{table}

\begin{table}[H]
\centering
\resizebox{\linewidth}{!}{
    \begin{tabular}{c|c|@{\extracolsep{14em}}ccccc@{}}
  \hline
  \hline
  Parameters & Levels & Median & Mean & SD & Lower HPD & Upper HPD \\ 
  \hline
  (Intercept) & -- & 5.325 & 5.343 & 0.102 & 5.173 & 5.542 \\\cline{1-2} 
  \multirow{8}{*}{age.car} & 0 & 0.023 & 0.021 & 0.080 & -0.109 & 0.163 \\ 
  & 1 & 0.084 & 0.081 & 0.080 & -0.066 & 0.215 \\ 
  & 2 & 0.194 & 0.192 & 0.080 & 0.051 & 0.330 \\ 
  & 3 & -0.036 & -0.040 & 0.080 & -0.187 & 0.094 \\ 
  & 4 & 0.021 & 0.020 & 0.080 & -0.126 & 0.155 \\ 
  & 5 & 0.072 & 0.068 & 0.080 & -0.074 & 0.209 \\ 
  & 6 & -0.022 & -0.025 & 0.081 & -0.168 & 0.116 \\ 
  & 7 & -0.072 & -0.078 & 0.083 & -0.224 & 0.057 \\ \cline{1-2}
  \multirow{1}{*}{risk} & S & -0.195 & -0.195 & 0.024 & -0.238 & -0.143 \\ \cline{1-2}
  \multirow{4}{*}{agec} & 2 & -0.135 & -0.135 & 0.032 & -0.200 & -0.075 \\ 
  & 3 & -0.479 & -0.480 & 0.031 & -0.540 & -0.421 \\ 
  & 4 & -0.626 & -0.626 & 0.036 & -0.696 & -0.559 \\ 
  & 5 & -0.694 & -0.694 & 0.059 & -0.814 & -0.581 \\ \cline{1-2}
  \multirow{2}{*}{gender} & F & 0.602 & 0.605 & 0.090 & 0.438 & 0.789 \\ 
  & M & 0.721 & 0.726 & 0.094 & 0.552 & 0.900 \\ \cline{1-2}
  \multirow{2}{*}{marital} & M & 1.527 & 1.623 & 0.286 & 1.197 & 2.152 \\ 
  & S & -0.354 & -0.339 & 0.258 & -0.763 & 0.222 \\  \cline{1-2}
  \multirow{7}{*}{deductible} & B & 1.371 & 1.366 & 0.222 & 0.921 & 1.815 \\ 
  & C & 0.147 & 0.088 & 0.253 & -0.353 & 0.495 \\ 
  & D & 0.535 & 0.465 & 0.247 & 0.010 & 0.829 \\ 
  & E & 0.718 & 0.637 & 0.243 & 0.207 & 1.010 \\ 
  & F & 0.680 & 0.598 & 0.243 & 0.178 & 0.979 \\ 
  & G & 0.680 & 0.599 & 0.240 & 0.188 & 0.970 \\ 
  & H & 0.766 & 0.791 & 0.331 & 0.143 & 1.436 \\   \cline{1-2}
  \multirow{4}{*}{genderMarital} & A & 0.226 & 0.215 & 0.223 & -0.240 & 0.623 \\ 
  & B & -2.005 & -2.065 & 0.257 & -2.551 & -1.664 \\ 
  & C & 0.281 & 0.275 & 0.273 & -0.278 & 0.720 \\ 
  & D & -2.023 & -2.107 & 0.299 & -2.604 & -1.680 \\   \hline
  \hline
\end{tabular}
}
\caption{\small Model coefficients for fixed effects in the mean model for M3.}\label{tab:mc-mean-spdglm-m3} 
\end{table}

\begin{table}[htb]
\centering 
\resizebox{\linewidth}{!}{
\begin{tabular}{c|c|@{\extracolsep{14em}}ccccc@{}}
  \hline
  \hline
  Parameters & Levels & Median & Mean & SD & Lower HPD & Upper HPD \\ 
  \hline
  (Intercept) & -- & 6.655 & 6.645 & 0.093 & 6.473 & 6.842 \\\cline{1-2} 
  \multirow{8}{*}{age.car} & 0 & -0.837 & -0.838 & 0.049 & -0.926 & -0.734 \\ 
  & 1 & -1.053 & -1.051 & 0.044 & -1.139 & -0.967 \\ 
  & 2 & -0.916 & -0.915 & 0.042 & -0.992 & -0.828 \\ 
  & 3 & -0.889 & -0.889 & 0.041 & -0.966 & -0.805 \\ 
  & 4 & -0.875 & -0.873 & 0.041 & -0.955 & -0.793 \\ 
  & 5 & -0.824 & -0.824 & 0.040 & -0.908 & -0.750 \\ 
  & 6 & -0.798 & -0.796 & 0.040 & -0.869 & -0.712 \\ 
  & 7 & -0.764 & -0.763 & 0.040 & -0.851 & -0.693 \\ \cline{1-2}
  \multirow{1}{*}{risk} & S & 0.105 & 0.104 & 0.015 & 0.075 & 0.134 \\ \cline{1-2}
  \multirow{4}{*}{agec} & 2 & -0.027 & -0.027 & 0.022 & -0.067 & 0.018 \\ 
  & 3 & -0.042 & -0.042 & 0.014 & -0.072 & -0.015 \\ 
  & 4 & -0.078 & -0.078 & 0.014 & -0.105 & -0.049 \\ 
  & 5 & -0.280 & -0.279 & 0.029 & -0.338 & -0.225 \\ \cline{1-2}
  \multirow{2}{*}{gender} & F & 0.004 & 0.003 & 0.038 & -0.071 & 0.077 \\ 
  & M & 0.005 & 0.005 & 0.043 & -0.076 & 0.089 \\ \cline{1-2}
  \multirow{2}{*}{marital} & M & -3.145 & -3.154 & 0.180 & -3.460 & -2.839 \\ 
  & S & -3.280 & -3.281 & 0.205 & -3.642 & -2.900 \\ \cline{1-2}
  \multirow{7}{*}{deductible} & B & -1.226 & -1.224 & 0.133 & -1.455 & -0.943 \\ 
  & C & -0.751 & -0.758 & 0.082 & -0.940 & -0.620 \\ 
  & D & -0.808 & -0.815 & 0.080 & -0.973 & -0.664 \\ 
  & E & -0.753 & -0.759 & 0.078 & -0.926 & -0.626 \\ 
  & F & -0.598 & -0.604 & 0.078 & -0.774 & -0.473 \\ 
  & G & -0.236 & -0.240 & 0.076 & -0.405 & -0.107 \\ 
  & H & 0.224 & 0.224 & 0.138 & -0.049 & 0.473 \\ \cline{1-2}
  \multirow{4}{*}{genderMarital} & A & 3.172 & 3.169 & 0.202 & 2.800 & 3.536 \\ 
  & B & 3.233 & 3.247 & 0.175 & 2.948 & 3.540 \\ 
  & C & 3.307 & 3.305 & 0.203 & 2.909 & 3.639 \\ 
  & D & 3.265 & 3.280 & 0.176 & 2.980 & 3.575 \\ \hline
  $\xi$ & -- & 1.671 & 1.671 & 0.004 & 1.664 & 1.679 \\
   \hline
   \hline
\end{tabular}
}
\caption{\small Model coefficients for fixed effects in the dispersion model for M3.}\label{tab:mc-disp-spdglm-m3}
\end{table}
   \singlespacing
    \bibliographystyle{apalike}
    \bibliography{bibliography}

\begin{thebibliography}{}

\bibitem[Abramowitz et~al., 1988]{abramowitz1988handbook}
Abramowitz, M., Stegun, I.~A., and Romer, R.~H. (1988).
\newblock Handbook of mathematical functions with formulas, graphs, and
  mathematical tables.

\bibitem[Agarwal et~al., 2002]{agarwal2002zero}
Agarwal, D.~K., Gelfand, A.~E., and Citron-Pousty, S. (2002).
\newblock Zero-inflated models with application to spatial count data.
\newblock {\em Environmental and Ecological statistics}, 9(4):341--355.

\bibitem[Akaike, 1998]{akaike1998information}
Akaike, H. (1998).
\newblock Information theory and an extension of the maximum likelihood
  principle.
\newblock In {\em Selected papers of hirotugu akaike}, pages 199--213.
  Springer.

\bibitem[Banerjee and Carlin, 2004]{banerjee2004parametric}
Banerjee, S. and Carlin, B.~P. (2004).
\newblock Parametric spatial cure rate models for interval-censored
  time-to-relapse data.
\newblock {\em Biometrics}, 60(1):268--275.

\bibitem[Banerjee et~al., 2014]{banerjee2014hierarchical}
Banerjee, S., Carlin, B.~P., and Gelfand, A.~E. (2014).
\newblock Hierarchical modeling and analysis for spatial data.

\bibitem[Berger et~al., 2001a]{berger2001objective1}
Berger, J.~O., De~Oliveira, V., and Sans{\'o}, B. (2001a).
\newblock Objective {B}ayesian analysis of spatially correlated data.
\newblock {\em Journal of the American Statistical Association},
  96(456):1361--1374.

\bibitem[Berger et~al., 2001b]{berger2001objective}
Berger, J.~O., Pericchi, L.~R., Ghosh, J., Samanta, T., De~Santis, F., Berger,
  J., and Pericchi, L. (2001b).
\newblock Objective {B}ayesian methods for model selection: {I}ntroduction and
  comparison.
\newblock {\em Lecture Notes-Monograph Series}, pages 135--207.

\bibitem[Berliner, 2000]{berliner2000hierarchical}
Berliner, M. (2000).
\newblock Hierarchical {B}ayesian modeling in the environmental sciences.
\newblock {\em AStA Advances in Statistical Analysis}, 2(84):141--153.

\bibitem[Best et~al., 2000]{best2000spatial}
Best, N.~G., Ickstadt, K., and Wolpert, R.~L. (2000).
\newblock Spatial {P}oisson regression for health and exposure data measured at
  disparate resolutions.
\newblock {\em Journal of the American statistical association},
  95(452):1076--1088.

\bibitem[Bradley et~al., 2018]{bradley2018computationally}
Bradley, J.~R., Holan, S.~H., and Wikle, C.~K. (2018).
\newblock Computationally efficient multivariate spatio-temporal models for
  high-dimensional count-valued data (with discussion).
\newblock {\em Bayesian Analysis}, 13(1):253--310.

\bibitem[Bradley et~al., 2020]{bradley2020bayesian}
Bradley, J.~R., Holan, S.~H., and Wikle, C.~K. (2020).
\newblock Bayesian hierarchical models with conjugate full-conditional
  distributions for dependent data from the natural exponential family.
\newblock {\em Journal of the American Statistical Association},
  115(532):2037--2052.

\bibitem[Carlin and Louis, 2008]{carlin2008bayesian}
Carlin, B.~P. and Louis, T.~A. (2008).
\newblock {\em Bayesian methods for data analysis}.
\newblock CRC press.

\bibitem[Carvalho et~al., 2010]{carvalho2010horseshoe}
Carvalho, C.~M., Polson, N.~G., and Scott, J.~G. (2010).
\newblock The horseshoe estimator for sparse signals.
\newblock {\em Biometrika}, 97(2):465--480.

\bibitem[Clark and Gelfand, 2006]{clark2006future}
Clark, J. and Gelfand, A. (2006).
\newblock A future for models and data in ecology.
\newblock {\em Trends in Ecology and Evolution}, 21:375--380.

\bibitem[Cressie, 2015]{cressie2015statistics}
Cressie, N. (2015).
\newblock {\em Statistics for spatial data}.
\newblock John Wiley \& Sons.

\bibitem[Dey et~al., 2000]{dey2000generalized}
Dey, D.~K., Ghosh, S.~K., and Mallick, B.~K. (2000).
\newblock {\em Generalized linear models: A {B}ayesian perspective}.
\newblock CRC Press.

\bibitem[Diggle et~al., 1998]{diggle1998model}
Diggle, P.~J., Tawn, J.~A., and Moyeed, R.~A. (1998).
\newblock Model-based geostatistics.
\newblock {\em Journal of the Royal Statistical Society: Series C (Applied
  Statistics)}, 47(3):299--350.

\bibitem[Dunn and Smyth, 2005]{dunn2005series}
Dunn, P.~K. and Smyth, G.~K. (2005).
\newblock Series evaluation of {T}weedie exponential dispersion model
  densities.
\newblock {\em Statistics and Computing}, 15(4):267--280.

\bibitem[Dunn and Smyth, 2008]{dunn2008evaluation}
Dunn, P.~K. and Smyth, G.~K. (2008).
\newblock Evaluation of {T}weedie exponential dispersion model densities by
  {F}ourier inversion.
\newblock {\em Statistics and Computing}, 18(1):73--86.

\bibitem[Eidsvik et~al., 2012]{eidsvik2012approximate}
Eidsvik, J., Finley, A.~O., Banerjee, S., and Rue, H. (2012).
\newblock Approximate {B}ayesian inference for large spatial datasets using
  predictive process models.
\newblock {\em Computational Statistics \& Data Analysis}, 56(6):1362--1380.

\bibitem[Fan and Liao, 2014]{fan2014endogeneity}
Fan, J. and Liao, Y. (2014).
\newblock Endogeneity in high dimensions.
\newblock {\em Annals of statistics}, 42(3):872.

\bibitem[Finley et~al., 2009]{finley2009hierarchical}
Finley, A.~O., Banerjee, S., and McRoberts, R.~E. (2009).
\newblock Hierarchical spatial models for predicting tree species assemblages
  across large domains.
\newblock {\em The annals of applied statistics}, 3(3):1052.

\bibitem[Gelfand et~al., 1996]{gelfand1996efficient}
Gelfand, A., Sahu, S., and Carlin, B. (1996).
\newblock Efficient {P}arametrizations for {G}eneralized {L}inear {M}ixed
  {M}odels.
\newblock {\em Bayesian Statistics 5: Proceedings of the Fifth Valencia
  International Meeting}, pages 165--180.

\bibitem[Gelfand, 2000]{gelfand2000modeling}
Gelfand, A.~E. (2000).
\newblock Modeling and inference for point-referenced binary spatial data.
\newblock {\em Generalized linear models: a Bayesian perspective}, page 373.

\bibitem[Gelfand and Banerjee, 2017]{gelfand2017bayesian}
Gelfand, A.~E. and Banerjee, S. (2017).
\newblock Bayesian modeling and analysis of geostatistical data.
\newblock {\em Annual review of statistics and its application}, 4:245--266.

\bibitem[Gelfand et~al., 1995]{gelfand1995efficient}
Gelfand, A.~E., Sahu, S.~K., and Carlin, B.~P. (1995).
\newblock Efficient parametrisations for normal linear mixed models.
\newblock {\em Biometrika}, 82(3):479--488.

\bibitem[Gelfand et~al., 2005]{gelfand2005modelling}
Gelfand, A.~E., Schmidt, A.~M., Wu, S., Silander~Jr, J.~A., Latimer, A., and
  Rebelo, A.~G. (2005).
\newblock Modelling species diversity through species level hierarchical
  modelling.
\newblock {\em Journal of the Royal Statistical Society: Series C (Applied
  Statistics)}, 54(1):1--20.

\bibitem[George and McCulloch, 1993]{george1993variable}
George, E.~I. and McCulloch, R.~E. (1993).
\newblock Variable selection via {G}ibbs sampling.
\newblock {\em Journal of the American Statistical Association},
  88(423):881--889.

\bibitem[Girolami and Calderhead, 2011]{girolami2011riemann}
Girolami, M. and Calderhead, B. (2011).
\newblock Riemann manifold {L}angevin and {H}amiltonian {M}onte {C}arlo
  methods.
\newblock {\em Journal of the Royal Statistical Society: Series B (Statistical
  Methodology)}, 73(2):123--214.

\bibitem[Halder et~al., 2021]{halder2022spatial}
Halder, A., Mohammed, S., Chen, K., and Dey, D.~K. (2021).
\newblock Spatial {T}weedie exponential dispersion models: an application to
  insurance rate-making.
\newblock {\em Scandinavian Actuarial Journal}, 2021(10):1017--1036.

\bibitem[Halder et~al., 2022]{halder2019spatial}
Halder, A., Mohammed, S., Chen, K., and Dey, D.~K. (2022).
\newblock Spatial risk estimation in {T}weedie double generalized linear
  models.
\newblock {\em Proceedings of International E-Conference on Mathematical and
  Statistical Sciences: A Selcuk Meeting}, 2022:62.

\bibitem[Heaton et~al., 2019]{heaton2019case}
Heaton, M.~J., Datta, A., Finley, A.~O., Furrer, R., Guinness, J., Guhaniyogi,
  R., Gerber, F., Gramacy, R.~B., Hammerling, D., Katzfuss, M., et~al. (2019).
\newblock A {C}ase {S}tudy {C}ompetition {A}mong {M}ethods {F}or {A}nalyzing
  {L}arge {S}patial {D}ata.
\newblock {\em Journal of Agricultural, Biological and Environmental
  Statistics}, 24(3):398--425.

\bibitem[Hoeting et~al., 1999]{hoeting1999bayesian}
Hoeting, J.~A., Madigan, D., Raftery, A.~E., and Volinsky, C.~T. (1999).
\newblock Bayesian model averaging: a tutorial (with comments by {M}. {C}lyde,
  {D}avid {D}raper and {EI} george, and a rejoinder by the authors.
\newblock {\em Statistical science}, 14(4):382--417.

\bibitem[Ishwaran and Rao, 2005]{ishwaran2005spike}
Ishwaran, H. and Rao, J.~S. (2005).
\newblock Spike and slab variable selection: frequentist and {B}ayesian
  strategies.
\newblock {\em The Annals of Statistics}, 33(2):730--773.

\bibitem[J{\o}rgensen, 1986]{jorgensen1986some}
J{\o}rgensen, B. (1986).
\newblock Some properties of exponential dispersion models.
\newblock {\em Scandinavian Journal of Statistics}, pages 187--197.

\bibitem[J{\o}rgensen, 1987]{jorgensen1987exponential}
J{\o}rgensen, B. (1987).
\newblock Exponential dispersion models.
\newblock {\em Journal of the Royal Statistical Society: Series B
  (Methodological)}, 49(2):127--145.

\bibitem[J{\o}rgensen, 1992]{jorgensen1992exponential}
J{\o}rgensen, B. (1992).
\newblock Exponential dispersion models and extensions: A review.
\newblock {\em International Statistical Review/Revue Internationale de
  Statistique}, pages 5--20.

\bibitem[Jorgensen, 1997]{jorgensen1997theory}
Jorgensen, B. (1997).
\newblock {\em The theory of dispersion models}.
\newblock CRC Press.

\bibitem[Kokonendji et~al., 2021]{bonat2021tweedie}
Kokonendji, C.~C., Bonat, W.~H., and Abid, R. (2021).
\newblock Tweedie regression models and its geometric sums for (semi‐)
  continuous data.
\newblock {\em Wiley Interdisciplinary Reviews: Computational Statistics},
  13(1):1496.

\bibitem[Lawson, 2018]{lawson2018bayesian}
Lawson, A.~B. (2018).
\newblock {\em Bayesian disease mapping: hierarchical modeling in spatial
  epidemiology}.
\newblock Chapman and Hall/CRC.

\bibitem[Lee and Nelder, 2006]{lee2006double}
Lee, Y. and Nelder, J.~A. (2006).
\newblock Double hierarchical generalized linear models (with discussion).
\newblock {\em Journal of the Royal Statistical Society: Series C (Applied
  Statistics)}, 55(2):139--185.

\bibitem[Li and Lin, 2010]{li2010bayesian}
Li, Q. and Lin, N. (2010).
\newblock The {B}ayesian elastic net.
\newblock {\em Bayesian analysis}, 5(1):151--170.

\bibitem[Liang et~al., 2008]{liang2008mixtures}
Liang, F., Paulo, R., Molina, G., Clyde, M.~A., and Berger, J.~O. (2008).
\newblock Mixtures of g-priors for {B}ayesian variable selection.
\newblock {\em Journal of the American Statistical Association},
  103(481):410--423.

\bibitem[Mallick et~al., 2022]{mallick2022differential}
Mallick, H., Chatterjee, S., Chowdhury, S., Chatterjee, S., Rahnavard, A., and
  Hicks, S.~C. (2022).
\newblock Differential expression of single-cell {RNA}-seq data using {T}weedie
  models.
\newblock {\em Statistics in medicine}, 41(18):3492--3510.

\bibitem[Martino et~al., 2011]{martino2011approximate}
Martino, S., Akerkar, R., and Rue, H. (2011).
\newblock Approximate {B}ayesian inference for survival models.
\newblock {\em Scandinavian Journal of Statistics}, 38(3):514--528.

\bibitem[Mat{\'e}rn, 2013]{matern2013spatial}
Mat{\'e}rn, B. (2013).
\newblock {\em Spatial variation}, volume~36.
\newblock Springer Science \& Business Media.

\bibitem[Mitchell and Beauchamp, 1988]{mitchell1988bayesian}
Mitchell, T.~J. and Beauchamp, J.~J. (1988).
\newblock Bayesian variable selection in linear regression.
\newblock {\em Journal of the american statistical association},
  83(404):1023--1032.

\bibitem[Mohammed et~al., 2021]{mohammed2021radiohead}
Mohammed, S., Bharath, K., Kurtek, S., Rao, A., and Baladandayuthapani, V.
  (2021).
\newblock {RADIOHEAD}: {R}adiogenomic analysis incorporating tumor
  heterogeneity in imaging through densities.
\newblock {\em The Annals of Applied Statistics}, 15(4):1808--1830.

\bibitem[Morris et~al., 2008]{morris2008bayesian}
Morris, J.~S., Brown, P.~J., Herrick, R.~C., Baggerly, K.~A., and Coombes,
  K.~R. (2008).
\newblock Bayesian analysis of mass spectrometry proteomic data using
  wavelet-based functional mixed models.
\newblock {\em Biometrics}, 64(2):479--489.

\bibitem[Nelder and Pregibon, 1987]{nelder1987extended}
Nelder, J.~A. and Pregibon, D. (1987).
\newblock An extended quasi-likelihood function.
\newblock {\em Biometrika}, 74(2):221--232.

\bibitem[Park and Casella, 2008]{park2008bayesian}
Park, T. and Casella, G. (2008).
\newblock The {B}ayesian lasso.
\newblock {\em Journal of the American Statistical Association},
  103(482):681--686.

\bibitem[Raftery et~al., 1997]{raftery1997bayesian}
Raftery, A.~E., Madigan, D., and Hoeting, J.~A. (1997).
\newblock Bayesian model averaging for linear regression models.
\newblock {\em Journal of the American Statistical Association},
  92(437):179--191.

\bibitem[Roberts and Stramer, 2002]{roberts2002langevin}
Roberts, G.~O. and Stramer, O. (2002).
\newblock Langevin diffusions and {M}etropolis-{H}astings algorithms.
\newblock {\em Methodology and computing in applied probability},
  4(4):337--357.

\bibitem[Shono, 2008]{shono2008application}
Shono, H. (2008).
\newblock Application of the {T}weedie distribution to zero-catch data in
  {CPUE} analysis.
\newblock {\em Fisheries Research}, 93(1-2):154--162.

\bibitem[Smyth, 1989]{smyth1989generalized}
Smyth, G.~K. (1989).
\newblock Generalized linear models with varying dispersion.
\newblock {\em Journal of the Royal Statistical Society: Series B
  (Methodological)}, 51(1):47--60.

\bibitem[Smyth and J{\o}rgensen, 2002]{smyth2002fitting}
Smyth, G.~K. and J{\o}rgensen, B. (2002).
\newblock Fitting {T}weedie's compound {P}oisson model to insurance claims
  data: dispersion modelling.
\newblock {\em ASTIN Bulletin: The Journal of the IAA}, 32(1):143--157.

\bibitem[Smyth and Verbyla, 1999]{smyth1999adjusted}
Smyth, G.~K. and Verbyla, A.~P. (1999).
\newblock Adjusted likelihood methods for modelling dispersion in generalized
  linear models.
\newblock {\em Environmetrics: The official journal of the International
  Environmetrics Society}, 10(6):695--709.

\bibitem[Swallow et~al., 2016]{swallow2016bayesian}
Swallow, B., Buckland, S.~T., King, R., and Toms, M.~P. (2016).
\newblock Bayesian hierarchical modelling of continuous non-negative
  longitudinal data with a spike at zero: An application to a study of birds
  visiting gardens in winter.
\newblock {\em Biometrical Journal}, 58(2):357--371.

\bibitem[Tweedie et~al., 1984]{tweedie1984index}
Tweedie, M.~C. et~al. (1984).
\newblock An index which distinguishes between some important exponential
  families.
\newblock In {\em Statistics: Applications and new directions: Proc. Indian
  statistical institute golden Jubilee International conference}, volume 579,
  page 6o4.

\bibitem[Verbyla, 1993]{verbyla1993modelling}
Verbyla, A.~P. (1993).
\newblock Modelling variance heterogeneity: residual maximum likelihood and
  diagnostics.
\newblock {\em Journal of the Royal Statistical Society: Series B
  (Methodological)}, 55(2):493--508.

\bibitem[Williams and Rasmussen, 2006]{williams2006gaussian}
Williams, C.~K. and Rasmussen, C.~E. (2006).
\newblock {\em Gaussian processes for machine learning}, volume~2.
\newblock MIT press Cambridge, MA.

\bibitem[Wolpert and Ickstadt, 1998]{wolpert1998poisson}
Wolpert, R.~L. and Ickstadt, K. (1998).
\newblock Poisson/gamma random field models for spatial statistics.
\newblock {\em Biometrika}, 85(2):251--267.

\bibitem[Yang et~al., 2018]{yang2018insurance}
Yang, Y., Qian, W., and Zou, H. (2018).
\newblock Insurance premium prediction via gradient tree-boosted {T}weedie
  compound poisson models.
\newblock {\em Journal of Business \& Economic Statistics}, 36(3):456--470.

\bibitem[Ye et~al., 2021]{ye2021comparisons}
Ye, T., Lachos, V.~H., Wang, X., and Dey, D.~K. (2021).
\newblock Comparisons of zero-augmented continuous regression models from a
  bayesian perspective.
\newblock {\em Statistics in Medicine}, 40(5):1073--1100.

\bibitem[Zeger and Karim, 1991]{zeger1991generalized}
Zeger, S.~L. and Karim, M.~R. (1991).
\newblock Generalized linear models with random effects; a {G}ibbs sampling
  approach.
\newblock {\em Journal of the American statistical association},
  86(413):79--86.

\bibitem[Zhang, 2002]{zhang2002estimation}
Zhang, H. (2002).
\newblock On estimation and prediction for spatial generalized linear mixed
  models.
\newblock {\em Biometrics}, 58(1):129--136.

\bibitem[Zhang, 2013]{zhang2013likelihood}
Zhang, Y. (2013).
\newblock Likelihood-based and {B}ayesian methods for {T}weedie compound
  {P}oisson linear mixed models.
\newblock {\em Statistics and Computing}, 23(6):743--757.

\bibitem[Zhou and Hanson, 2015]{zhou2015bayesian}
Zhou, H. and Hanson, T. (2015).
\newblock Bayesian spatial survival models.
\newblock {\em Nonparametric Bayesian Inference in Biostatistics}, pages
  215--246.

\end{thebibliography}
    \end{document}